\documentclass[trackchanges]{aastex631}
\usepackage{amsmath}
\usepackage{gensymb}

\shorttitle{Source positions for quad-band VLBI observations}
\shortauthors{Xu \& Charlot}

\graphicspath{{./}{figures/}}

\begin{document}

\title{Variations of absolute source positions determined from quad-band VLBI 
observations}

\email{Corresponding to: minghui.xu@gfz-potsdam.de}

\author[0000-0001-9602-9489]{Ming Hui Xu}
\affiliation{GFZ Helmholtz Centre for Geosciences, Telegrafenberg, 14476,
Potsdam, Germany}
\author[0000-0002-9142-716X]{Patrick Charlot}
\affiliation{Laboratoire d’astrophysique de Bordeaux, Univ. Bordeaux,
CNRS, B18N, Allée Geoffroy Saint-Hilaire, 33615 Pessac, France}
%



\begin{abstract}

Active Galactic Nuclei (AGNs) observed with the technique of very
long baseline interferometry (VLBI) are used as fiducial references on the sky
to precisely measure the shape and orientation of the Earth.
Their positions form a celestial reference frame that plays an
important role in both astronomy and geodesy. This study investigates the
accuracy and stability of the positions of the AGNs that are measured by 
simultaneous VLBI observations at 3.3, 5.5,
6.6, and 10.5\,GHz. Based on position time series from dedicated geodetic
solutions, we characterize the observed source position variations and
identify the possible factors causing such variations. We find that the
primary contributor is source structure for sources
above 20$\degree$ declination while the sensitivity of the observations
to the declination coordinate predominates for sources below
20$\degree$ declination. The position time series are further explored to
derive more realistic uncertainties for the quad-band positions. Significant
position offsets with respect to the positions at 2.2/8.6\,GHz are found
for 15\% of the sources. For 6\% of the sources, the offsets are larger
than 0.8\,milli-arcseconds. Source
structure may be divided into two parts: the invisible structure (within
the beam size) and the visible structure (on larger scales). The latter causes
closure delays enlarging post-fit delay residuals in geodetic solutions whereas
the former causes source position changes. Such position changes will contribute
significantly to the offsets between radio and optical positions. Overall, this
work highlights the necessity to have a specific
quad-band catalog for processing operational quad-band observations.


\end{abstract}



\section{Introduction} \label{sec:intro}

The new generation geodetic very-long-baseline interferometry (VLBI) system,
named VLBI Global Observing System (VGOS) and under development since the
mid-2000
\citep[see, e.g.,][]{2009vlbi.rept....1P,2018RaSc...53.1269N}, has started to
make regular observations to measure station positions of globally distributed
antennas and Earth Orientation Parameters (EOPs) since 2019. With a
different technical design than the legacy geodetic VLBI system used since 1978,
namely a much wider frequency coverage and faster slewing
rates allowed by smaller antennas, VGOS attempts to improve the
accuracy and stability of the geodetic products, that is, station
position/velocity and EOPs \citep{2016JGRB..121.6109A,2019JGeod..93..621B}.
The VGOS network is still expanding to new stations worldwide for the
purpose of supporting geodesy and astrometry through the next several decades,
like the legacy VLBI system has done for the past several decades.
One of the key elements for this new system,
necessary to achieve its goal, is the availability of an astrometric catalog
with accurate radio source positions as anchors on the sky and the high
stability of these positions
over time. The radio sources observed by astrometric/geodetic VLBI
are Active Galactic Nuclei (AGNs). Their radio emission is
dominated by
synchrotron radiation from relativistic jets of plasma launched in the vicinity
of supermassive black holes, and shows extended structure that usually
consists of a bright featureless component (core)
and multiple components along a single-sided jet. The angular structure of the
radio emission may vary both with time and frequency. Moreover, the sources
observed by VGOS are all very bright, making their jets more
likely to be detected. The variation of source structure with time
can lead to instabilities in the source positions. The variation with
frequency can lead to position differences with respect to those measured by
the legacy geodetic VLBI system because different
frequency bands are used by the VGOS and legacy systems. VGOS simultaneously 
observes in four bands, each currently 512\,MHz wide, centered at
3.3,\,5.5,\,6.6,\,and 10.5\,GHz, whereas the legacy system observes in the S
and X bands (2.2 and 8.6\,GHz). In the following, we will refer to the source
positions determined from
quad-band VLBI observations as quad-band positions and those from legacy VLBI
observations as S/X positions. From the geodetic point of view, it is important 
and interesting to investigate the time variability of
the source positions. To date, investigations of such
variability have
been focused on the S/X positions \citep[see, e.g.,
][]{2017JGeod..91..755K,2020A&A...634A..28L,2021A&A...648A.125G,
2024A&A...684A..93L,2024ApJS..274...28C}.

In astronomy, the fundamental celestial reference frame (CRF) 
has been realized by using the observations from the legacy geodetic VLBI 
system since 1998
\citep{1998AJ....116..516M,2015AJ....150...58F,2020A&A...644A.159C}. Recently,
the European Space
Agency mission Gaia\footnote{\url{https://sci.esa.int/web/gaia}} 
\citep{2016A&A...595A...2G} has established the CRF at 
optical wavelengths with a comparable accuracy level 
\citep{2022A&A...667A.148G}. The median difference of radio and optical source 
positions for more than 2000 common sources with optical $G$ magnitudes up to
21 is about 0.5\,milli-arcseconds (mas),
while when restricting the comparison to sources with optical $G$ magnitudes
$<$ 18.0 (which have higher position accuracy in the optical) it is on the
order of 0.3\,mas \citep{2021A&A...647A.189X}. The vectors joining the
radio and optical positions are found to be parallel to the directions of the
radio jets for the sources that have significant position offsets, about
10--20\% of the common sources
\citep{2017A&A...598L...1K,2017MNRAS.467L..71P,2019ApJ...871..143P,
2021A&A...647A.189X}. The magnitude of the offsets is on the mas level for
these sources, and it is suggested that such large differences are caused by
optical jets, meaning that the
optical jet emission observed by Gaia and the radio jet emission observed by
VLBI originate in different regions of the objects. For the
majority ($\sim$80\%) of the sources, however, the position
differences are on the sub-mas order, on which scales factors like radio source
structure and VLBI position uncertainties play a role. Separation of the
various phenomena that can cause optical--radio
position differences \citep{2021A&A...652A..87L,2024A&A...684A.202L} and
investigating the potential
contributions of astrophysical processes such as the opacity effect
\citep{2008A&A...483..759K,2011A&A...532A..38S} is not straightforward. In this
respect, accumulation of data to monitor the source positions and improving the
position accuracy will be essential to further understand the various
contributions.

Besides the construction of the CRF other astronomical applications, among which
those using differential (or relative) astrometry, require
accurate radio source positions. The VLBI phase referencing technique can
measure the relative positions at the micro-arcsecond level between a nearby
pair of target and calibrator \citep{2014ARA&A..52..339R,2020A&ARv..28....6R}.
For instance, by relying on the absolute positions of the calibrators ---
AGNs ---
phase referencing observations permit the determination of the absolute
positions of the radio stars in our Galaxy as well as their proper motions and
parallaxes
\citep[e.g.,][]{2023MNRAS.524.5357C,2023A&A...676A..11L,2024MNRAS.529.2062Z},
which can be used to link the Gaia frame to the radio CRF
\citep{2020A&A...633A...1L,2024A&A...689..A134}. However, the
absolute positions of the AGNs are affected by their radio structure --- the
positions of the AGNs in the CRF catalog may not agree with the location of the
core or the brightest feature in a radio image --- and the errors are
propagated into the radio positions of the stars, limiting the optical--radio
frame tie. Recently, \citet{2024A&A...689..A134} have identified that the
uncertainties of the positions of the AGNs are one of the major error sources in
determining the astrometric parameters of the stars via phase referencing.

The aim of this study is three-fold: (1) investigating the
stability of the quad-band positions over time, (2) studying the factors
causing position variations, and (3) evaluating the differences between
the quad-band and S/X positions.
Recently, \citet{2024AJ....168...76P} evaluated the accuracy of quad-band 
positions by comprehensively analyzing VLBI observations with a frequency 
range from 2 to 43 
GHz and comparing the position differences across this frequency range. In this
study, we will use a different method, which involves deriving source position
time series based solely on VGOS observations. This approach will allow
us to
evaluate the position stability of each individual source,
identify the various factors causing the position variability, and derive
realistic position uncertainties, which will be essential for comparing with the
positions obtained from legacy geodetic VLBI observations and establishing a
VGOS CRF.

\section{Data and data analysis} \label{sec:data}

The quad-band VLBI data used in this study are the 24-hour experiments that
are coordinated by the International VLBI Service for Geodesy and Astrometry
\citep[IVS, ][]{2012JGeo...61...68S}. These data are publicly available from
the IVS data
archive\footnote{\url{https://cddis.nasa.gov/archive/vlbi/ivsdata/vgosdb/}}.
The primary VLBI observable for geodetic purposes is the group delay
(or simply delay for short), which is the difference in the arrival
times of a wavefront from a radio source between the two antennas of a
baseline. By correlating the two signals recorded on a
baseline, the visibility phase and visibility amplitude are obtained, and
the group delay is then determined as the partial derivative of the visibility
phase with respect to frequency. In VGOS the group delay is fitted
from the visibility phases over the four bands simultaneously. In a typical
quad-band VLBI experiment (session),
group delay observations are produced by a network of about ten antennas for
tens to hundreds of radio sources during a period of 24 hours. Overall, each
session includes an average of about 9000 observations. These group delays are
analyzed in geodetic solutions to derive, for instance, the absolute
positions of radio sources. We analyzed 177 quad-band
experiments conducted with a total of 13 antennas from 03 December 2017
to 26 June 2024 using NASA VLBI data analysis software
\emph{pSolve}\footnote{\url{https://astrogeo.org/psolve/}} --- originated
from Calc/Solve software. The theoretical delay model applied follows the IERS
2010 conventions \citep{2010ITN....36....1P} with the Galactic aberration
effect (using the aberration constant of 5.8 $\mu$as/yr) taken into
account. Details about the delay model in \emph{pSolve} software can be found in
\citet{2011AJ....142...35P} and \citet{2023AJ....165..183P}. The group delay
observations were analyzed in two steps: (1) session-wise data editing and (2)
processing observations from multiple sessions in a combined (global) solution.
These two steps are described in the following. Additionally, the quad-band
data were analyzed for astronomical imaging by using the phase and
amplitude observables. The processing for this imaging work is also
described briefly below.

\subsection{Session-wise data editing} \label{sec:session-wised}
In the data editing step, each session is processed 
independently for detecting clock jumps, flagging outliers, adjusting the 
constraints for clocks and atmospheric parameters, and determining the 
baseline-based additive weights. The additive weights are introduced to adjust
the group delay uncertainties so that they match
the post-fit residual level, that is, to achieve a delay residual $\chi^{2}$ per degree of
freedom being unity. It is not necessary in this initial stage to estimate
source positions, especially for the routine experiments where the aim is
to measure EOPs. In the quad-band VLBI databases from the IVS data archive,
source positions were estimated for all sources (with more than 3 usable delay
observations) in the sessions before vr2202 (17 March 2022) but were not
estimated for any of sources after that. One should note that a VLBI database
containing the measurements from a specific session
has a different set of usable group
delays depending on whether the source positions were estimated or
not because the outlier detection is affected if the source positions have large
errors. For example, an error of 0.5\,mas in a source position can cause delay
residuals of about 65\,picoseconds (ps) on observations for an 8000-km-long
baseline, leading to such observations being excluded as outliers in a solution
with a weighted root-mean-square (wrms) delay residual of 20\,ps and a cut-off
threshold of 3$\sigma$. Since the quad-band data
analysis relies on the S/X positions as a priori --- we do not
have a quad-band source position catalog available yet --- one would
expect large delay residuals for some sources. To recover these potentially
good measurements with good robustness, additional rounds of solutions were
performed.

After obtaining a complete solution of one session without estimating any source 
position, we examined two statistical values: the source-based delay residual
$\chi^2$ per degree of freedom and the number of outliers per source. The 
estimation of source positions was then enabled for the sources that either
have the residual $\chi^{2}$ per degree of freedom larger than 1.2 (20\% higher
than the overall $\chi^{2}$ of all sources) or have more than ten outliers. 
After running a solution with this setting, the applied parameterization (i.e.,
the estimation of source positions) was kept only for the sources that show
position shifts three times larger than their formal errors. After that, the
outlier restoration/detection and the adjustment of weights were performed in an
iterative way until the
solution converged. It is common to have 10 -- 20 radio sources with their
positions estimated in a session and to recover 100 -- 300 observations through
the iterative process. This number of
recovered observations corresponds to a few percents of the total number of
observations in each session and is significant for they are recovered
from a rather complete solution with a typical outlier fraction of only 1 --
5\%.

\subsection{Global solutions}
Geodetic and astrometric sessions can be combined
in a global solution to solve for parameters like station and source positions
globally (i.e., estimating them as best fits over all sessions) or for
parameters like clocks locally (i.e., estimating them over a single
session or intra-session time period). In the global solution of the 177
quad-band sessions investigated here, we used the following
parametrization:
\begin{itemize}
\item{global parameters (constant over the time span of all sessions)}
  \begin{enumerate}
  \item{source positions (except for the sources for which position is estimated
session-wise)}
  \item{station positions}
  \item{station velocity of station \texttt{ISHIOKA}}
  \end{enumerate}
  \item{local parameters (over the time intervals no longer
than 24 hours, specified below)}
  \begin{enumerate}
  \item{station clocks with a one-hour interval, except in the five sessions
from 22 March 2023 to 14 June 2023 for which a 10-minute interval was used to
account for the rapid clock variations at stations \texttt{HOBART12} and 
\texttt{KATH12M}}
  \item{zenith tropospheric delays with a 20-minute interval}
  \item{atmospheric gradients with a 6-hour interval}
  \item{polar motion and UT1 as well as their rates (once per session)}
  \item{nutation (once per session), only for sessions with a network of more
than six stations}
  \item{source positions for selected radio sources (discussed in following
sections, once per session)}
  \end{enumerate}
  \end{itemize}

The a priori source positions are from the S/X catalog of the third realization 
of the International Celestial Reference Frame 
\citep[ICRF3, ][]{2020A&A...644A.159C}. The a priori station
positions for four stations, \texttt{KOKEE12M}, \texttt{ONSA13NE}, 
\texttt{ONSA13SW}, and \texttt{WETTZ13S} were
linked to the positions of their co-located legacy stations in the latest
realization of the International Terrestrial Reference Frame \citep[ITRF2020,
][]{2023JGeod..97...47A}, rather
than taking their ITRF2020 positions directly. This
link was done by using the position ties, reported in
\citet{2023JGRB..12825198X}, between the VGOS stations and the legacy stations
that were obtained internally with VLBI observations. The positions of the
remaining 9 VGOS stations were taken from the ITRF2020. The station
velocities used are the ITRF2020 velocities of their respective co-located
legacy stations. For station \texttt{MACGO12M}, which is not part of the
ITRF2020 and does not have a legacy station at the site, the velocity of
station \texttt{FD-VLBA} in the ITRF2020 was used (the distance between them
is 9 km). The velocity of station
\texttt{ISHIOKA} was estimated because it shows a significant discrepancy with
respect to the ITRF2020. Five stations, the four stations
mentioned above that have highly-precise position ties and station
\texttt{WESTFORD} (same 
antenna for the VGOS and the legacy system), were used for the datum
definition of the terrestrial reference frame
(TRF), meaning that these stations define fiducial references on the Earth's
surface. To this end, station coordinates were estimated subject to
no-net-translation and no-net-rotation constraints applied with respect to
the a priori coordinates for these datum stations. This specific TRF
configuration setting for processing quad-band observations is not further
discussed in the present paper but will be explained in a separate publication
on VGOS station positions (Xu et al., in preparation). One reason for omitting
the discussion here is that the impact of the TRF datum on source
position time series is minimal.

Forty-six radio sources served for the datum definition of the CRF, i.e., the
no-net-rotation constraints were
applied with respect to their ICRF3 positions. In the solution, the positions of
seven sources were estimated per session as local parameters.
The selection of the 46 datum sources --- 27 of them are ICRF3
defining sources --- and the seven local sources is based on
the analysis of their position stability and is explained
in the following sections.

Fifty-seven radio sources that have fewer than 20 group
delays accumulated over the 177 sessions and seven radio sources that have fewer
than 30 accumulated group delays and no more than six group delays in any single
session were excluded. In total, 1,633,781 group delays on 377 radio sources
were included in the global solutions.

\subsection{Deriving source position time series}\label{obtain_time_series}
Source position relates to the location of a radio source on a celestial sphere
and consists of two coordinates, right ascension and declination. To be
precise, in this study, we will refer to it as the position estimate obtained
from analyzing radio or optical observations at some given frequency bands.
In VLBI, the astrometric position measured for an object tends to be
close to the location of the peak component of its radio emission because the
observations are made with interferometry. This contrasts to measuring the
brightness centroid of the emission, as is done for example in the optical with
Gaia. Since source structure evolves with both time and frequency, the
precise connection between the source position (as measured by VLBI
observations of a certain time period) and radio emission (as shown by an image
at one epoch), however, remains complex, in particular, down to the
level of tens of micro-arcseconds. Nevertheless, an important property
for geodesy and astronomy is the stability of the source positions over time.
One should note that the quad-band source positions are derived from the
group delays fitted from four band visibilities simultaneously,
rather than from separately obtaining the position in each of four bands (by
processing independently the group delays at each band) and
averaging. For the comparison of the quad-band positions, the astrometric
positions at S/X, K, and X/Ka bands used in the study are taken from the ICRF3
\citep{2020A&A...644A.159C}.

A crucial element when deriving source position time
series is the availability of a stable datum for the CRF such that the
intrinsic variation in source position for a given source can be properly
measured. This is best achieved in global solutions where
source positions and station positions are
treated as global parameters across different epochs. This allows for the same
group of sources to be used for the CRF datum and the same group of stations for
the TRF datum across all 177 sessions. In comparison, session-wise solutions are
more sensitive to the configuration of the observations in each session
(station network, sources, and so on) and the CRF/TRF datums available in these
solutions have to vary from session to session. Furthermore, to maintain a
stable CRF datum for different sources, we performed one dedicated global
solution for each radio source, in which the coordinates of this radio source
were estimated from each session as local parameters. In total, 190
global solutions were run one by one for the 190 radio sources that have been 
observed in more than ten sessions. The wrms delay residuals from
these 190 solutions are in the range of 17.1 to 17.4 ps, and the $\chi^{2}$
per degree of freedom is in the range of 0.83 to 0.84. Details on the source
position variations and the categorization are discussed in
Sect.\,\ref{sec:variations}.

\subsection{Astronomical imaging}
In order to understand the observed variations in the astrometric positions, we
processed the quad-band visibilities (phases and amplitudes) to derive
structure images. The images presented in this
study were obtained through three steps: (1) imaging quad-band VLBI data
by using closure (visibility) phases and closure (visibility) amplitudes only
\citep[see,][]{2018ApJ...857...23C,2021JGRB..12621238X} because of the lack of
amplitude
calibration in quad-band VLBI observations; (2) calibrating the visibility
data based on the obtained images; and (3) performing model fitting using 
\emph{difmap} \citep{1994BAAS...26..987S} to model
the structure with a small number of Gaussian components. We will refer to the
images derived with this approach as closure-only images.

\section{Variations in source positions} \label{sec:variations}

For further analysis of the position time series, session-wise
position estimates with uncertainties larger than 1\,mas or with a
strong correlation between the two coordinates, larger than 0.7, were excluded.
To get insights into the overall variation level, the
standard deviations of the position time series with respect to the mean
positions were calculated. They have a mean value of 0.14\,mas
for RA$^{*}$\footnote{RA$^{*}$
refers to the angular distance in the right ascension direction, i.e.,
RA*cos(Dec).} and
0.25\,mas for Dec over the 190 sources. For the sources with declinations higher
than 20$\degree$, the averaged standard deviations are 0.10\,mas for RA$^{*}$
and 0.14\,mas for Dec. By examining the 190 position time series, we could
identify three different types of variations in the
quad-band source positions: case of stable or quasi-stable positions,
continuous motion in one direction, and back-and-forth motion along one
direction.

\subsection{Case of stable or quasi-stable positions}

\begin{figure}[ht!]
\plotone{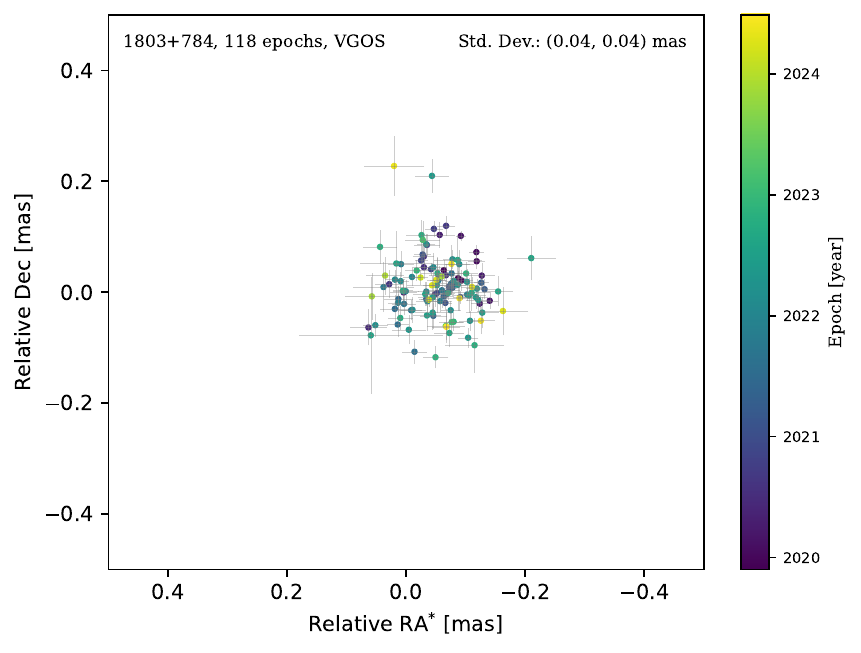}
\caption{Distribution of positions of source 1803+784 at 118 epochs determined 
from a global solution of quad-band VLBI observations. The positions are
relative to the S/X position of 1803+784 in the ICRF3. Color coding depicts the
epoch of each of these 118 VLBI sessions. The standard deviation of the
position estimates with respect to the weighted mean is 0.04\,mas for both
coordinates, as written in the upper-right corner. Position estimates with a
strong correlation ($>$ 0.7) between right ascension and declination are not
shown as they rely on only a few observations and have very large uncertainties.
\label{fig:2d_1803+784}}
\end{figure}

The position time series of source 1803+784, obtained from the global solution,
is shown in Fig. \ref{fig:2d_1803+784} as an example of the cases
where the position remains stable within a few tens of micro-arcseconds $\mu$as
over the entire data span (i.e. 118 sessions spanning 4.5 years). The
median number of group delays in these 118 sessions is 698. This high number of
observations per source in a single experiment is warranted by the fast slewing
rates of VGOS
antennas and newly developed approaches in scheduling
\citep{2023JGeod..97...17S,2024arXiv240713323S}. The high
stability that is seen in Fig.\,\ref{fig:2d_1803+784} also partly
results from the low measurement noise level in the quad-band observations,
which is one order of magnitude lower than that in the S/X observations
\citep{2018JGRB..12310162A,2021JGeod..95...51X}. The quad-band observations thus
provide more precise position time series than the S/X observations. The median
uncertainty is 0.02\,mas for both of RA$^{*}$ and declination while the
reported uncertainty in the ICRF3 is 0.03\,mas. The
standard deviation with respect to its weighted mean is 0.04\,mas for both
coordinates. The weighted mean position (RA$^{*}$, Dec) with respect to the S/X
position is ($-$0.040, 0.011) $\pm$ (0.004, 0.004)\,mas. This small but
statistically significant offset in right ascension (also seen in
Fig.\,\ref{fig:2d_1803+784}) agrees with the
fact that the source has an extended jet along the right ascension direction
\citep[][see Fig. 1 therein]{2024JGeod..98...38K}. Another feature to
consider in the distribution of position estimates is whether they show
systematic trends over time. In the case of 1803+784, there might be a small
trend between the year 2020 and after, but apart from that, the position is
essentially stable. Based on this study, we conclude
that the uncertainty of the source position estimated from a single session in
the global solution performed for this source is comparable to, or only
slightly lower than, the position stability itself, where the position
stability is defined by
the wrms of the residuals of the time series relative to the mean (or relative
to an empirical model that attempts to describe the position variations).
The results for 1803+784 highlight the capability of quad-band
observations to determine source positions at the level of 0.02 to 0.04 mas
per epoch for a single session under good observing conditions. This
justifies the use of the formal errors from the global
solutions to explore source position time series, i.e.,
they represent the level of the measurement noise and can be used to assess the
position stability and evaluate other systematic effects.
Several other frequently observed sources, such as 0602+673, 
0613+570, 0716+714, and 0814+425, were found to have similar position
stability. Two-dimensional plots such as that in Fig.\,\ref{fig:2d_1803+784}
have been generated for 96 sources, including these
four sources, and are available in the supplementary material of the paper.

The case of 1803+784 reveals that a source can have a stable position despite
being not point-like. In general, the extended structure on scales of several
milli-arcseconds, such as that is seen for 1804+784, cannot draw the estimated
source positions significantly away from the location of the brightest
component (identified as the core in most cases). Instead, it only affects
the astrometric positions at the level of a few tens of micro-arcseconds. This
is because the radio emission is dominated by that from the core, with which
the emission from the jet components does not ``beat'' in a coherent way
due to their separations being greater than the interferometer beam size.
Following \citet{2010ivs..conf....8P}, we refer to such extended structure
(away from the core component) as ``visible'' structure.
The fact that the visible
structure only changes source positions at the level of less than 0.1\,mas
has been shown from simulation studies \citep{2016MNRAS.455..343P} and
confirmed by \citet{r:gaia3} who applied source structure corrections based on
images, produced from the Monitoring Of Jets in Active galactic nuclei with
VLBA Experiments
project\footnote{\url{https://www.cv.nrao.edu/MOJAVE/index.html}}
(MOJAVE) observations, in geodetic data analysis. However, it is important to 
note that the visible structure introduces significant delay errors, which
manifest as closure delays in baseline-based
fringe fitting used in geodetic data processing. These closure
delays will affect the geodetic solutions and bias the other geodetic parameters
\citep{2016AJ....152..151X,2017JGeod..91..767X,2018JGRB..12310162A}.

\subsection{Continuous motion in one direction}\label{one_motion}

\begin{figure}[ht!]
\plotone{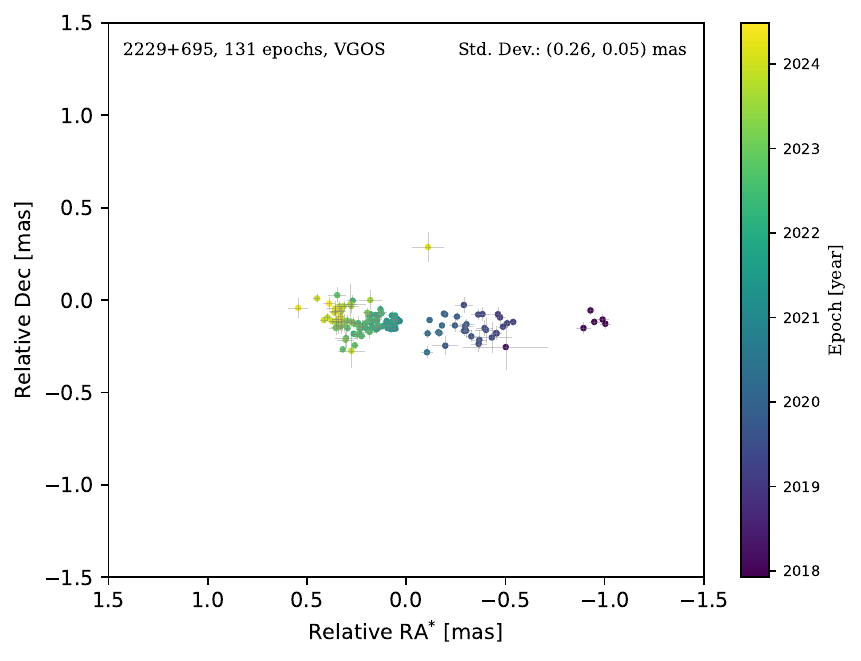}
\caption{Distribution of the quad-band positions of source 2229+695, with 
respect to its S/X position in the ICRF3, at 131 epochs. The standard deviation
of the
positions with respect to the 
weighted mean value is shown in the upper-right corner. \label{fig:2d_2229+695}}
\end{figure}

\begin{figure}[ht!]
\gridline{
          \fig{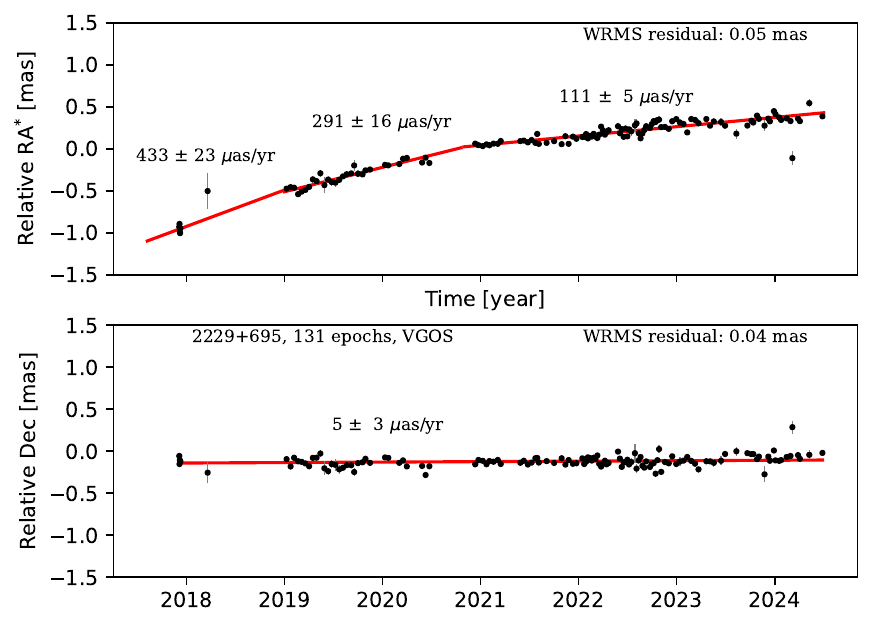}{0.5\textwidth}{}
          }
\caption{Time series of the quad-band positions of source 2229+695. Three 
linear segments were fitted to the RA$^{*}$ coordinate, while one segment was
fitted to the Dec
coordinate on the entire time span. The fitted segments are shown as
red lines. The weighted rms residual with respect to the fitted segments that
model the source motion is
0.05\,mas for RA$^{*}$ and 0.04\,mas for declination. \label{fig:ts_2229+695}}
\end{figure}

The position estimates of source 2229+695 are shown as a two-dimensional
distribution in Fig.\,\ref{fig:2d_2229+695} and as right ascension and
declination time series in Fig.\,\ref{fig:ts_2229+695}. Between 2018 and 2024,
the position of 2229+695 moved continuously towards the East with a
displacement of 1.7\,mas. The observed
variations with time may be modeled as three linear segments for RA$^{*}$ and
one linear segment for Dec. The segments in RA$^{*}$ correspond to a
velocity of 433\,$\pm$\,23\,$\mu$as/yr at the beginning,
291\,$\pm$\,16\,$\mu$as/yr from
2019 to 2021, and 111\,$\pm$\,5\,$\mu$as/yr after 2021, while that in Dec.
corresponds to a non-significant velocity of 5$\pm$3 $\mu$as/yr.
These segments are shown in Fig.\,\ref{fig:ts_2229+695} as red curves. The
position time series has a standard deviation, with respect to the weighted
mean, of 0.26\,mas for RA$^{*}$ and 0.05\,mas for Dec, whereas the wrms of the
position residuals with respect to the modeled segments is only 0.05\,mas for
RA$^{*}$ and 0.04\,mas for Dec, comparable to the position
stability of source 1803+784. 

\begin{figure}[ht!]
\gridline{
          \fig{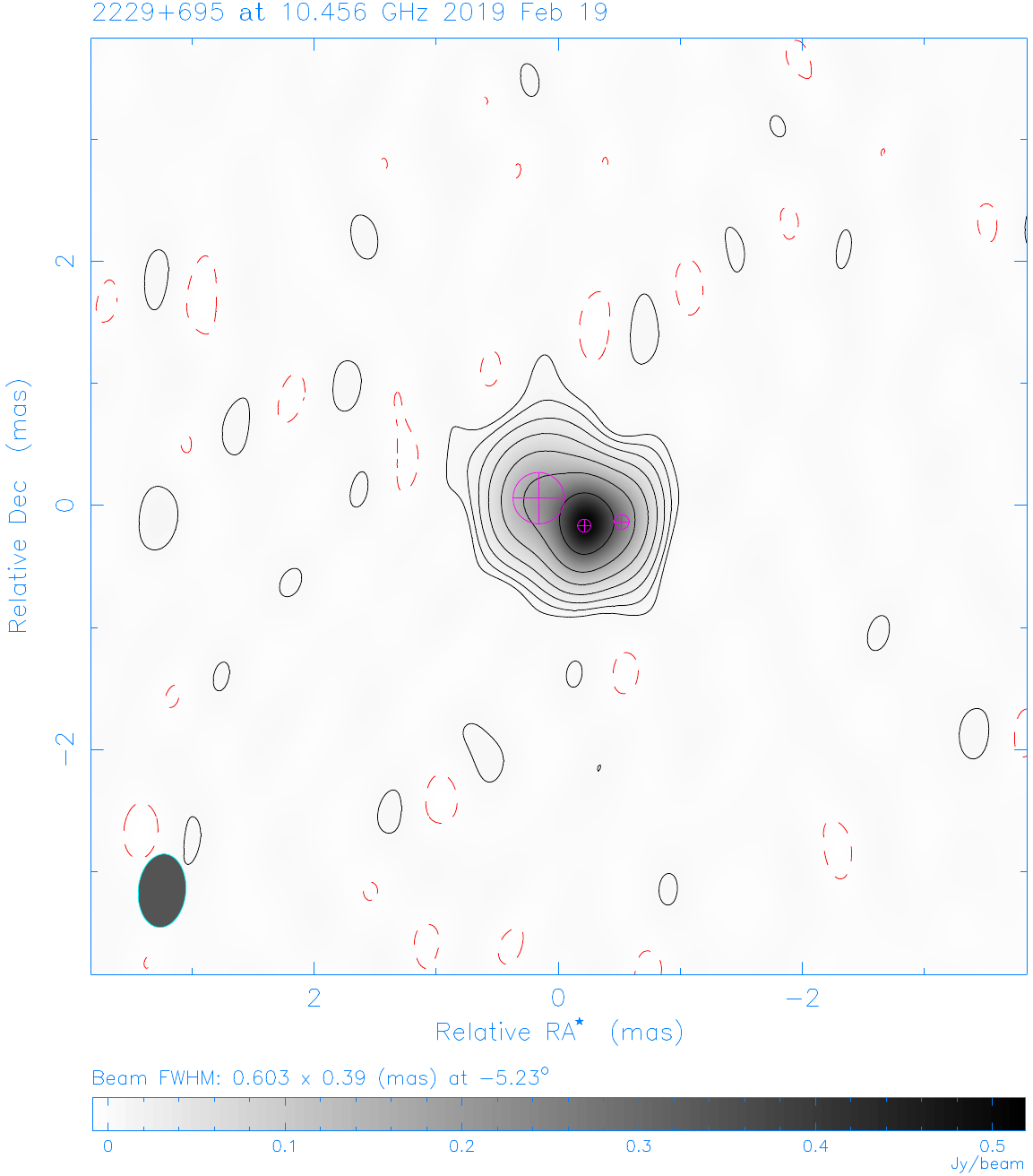}{0.25\textwidth}{}
          \fig{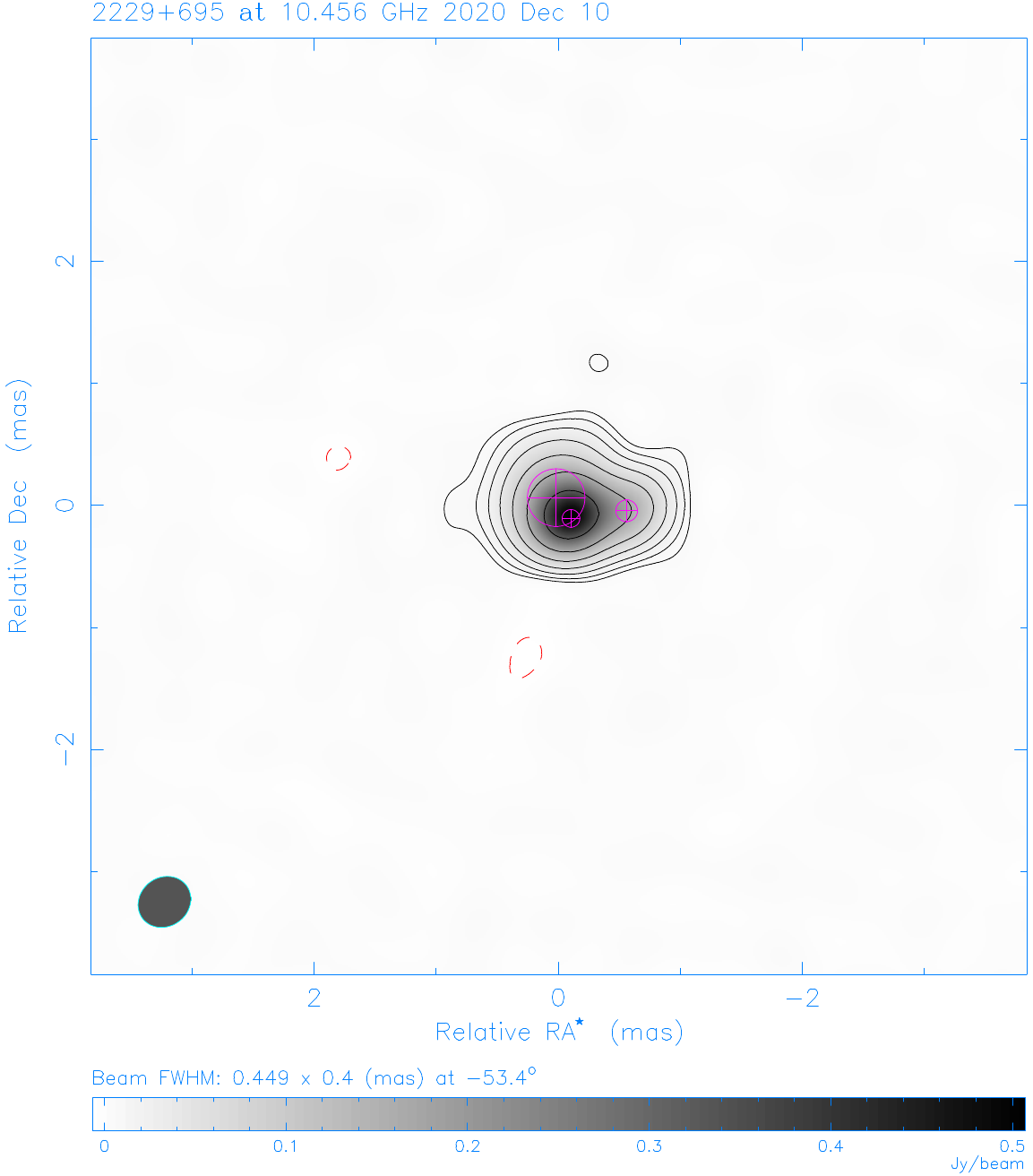}{0.25\textwidth}{}
          \fig{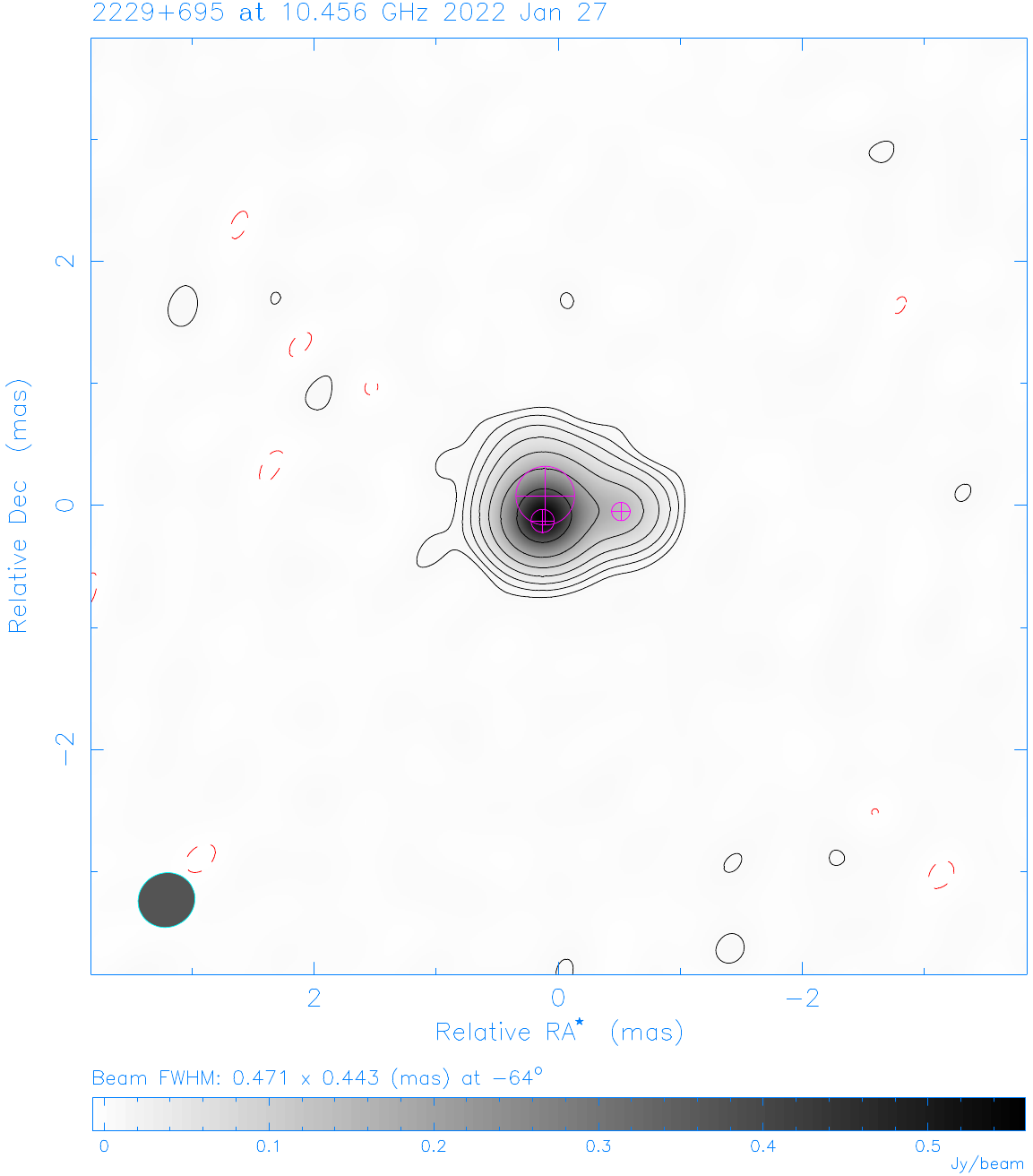}{0.25\textwidth}{}
          \fig{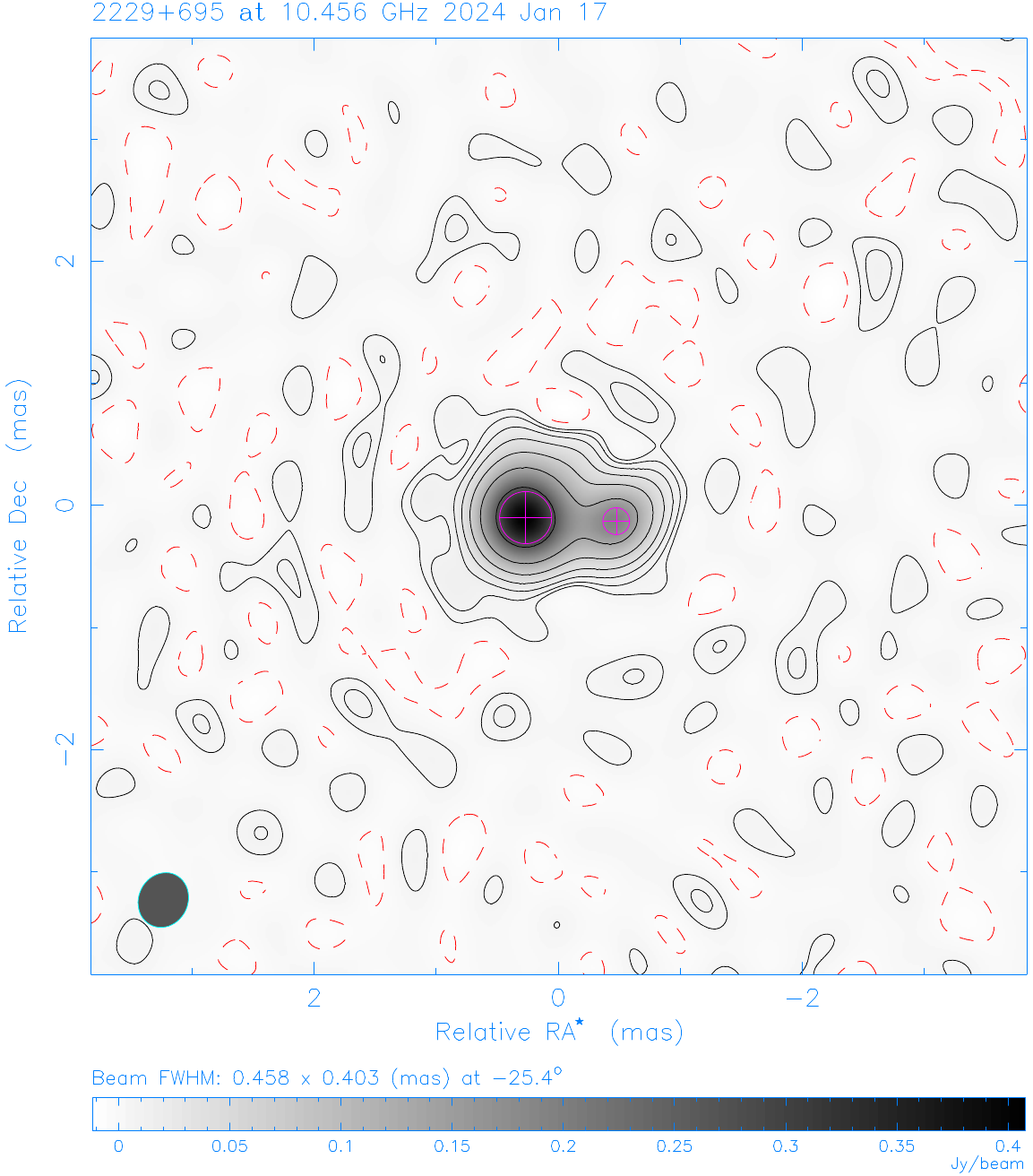}{0.25\textwidth}{}
          }
\caption{Closure-only images of source 2229+695 at 10.5\,GHz for four epochs:
19 February
2019 (first), 
10 December 2020 (second), 27 January 2022 (third), and 17 January 2024
(fourth). These images were obtained from quad-band VLBI sessions vt9050,
vo0345, vo2027, and vo4017, respectively. The circular Gaussian components 
determined from model fitting are shown as magenta circles. The
angular
distance in RA$^{*}$ between the peak component and the Western component was
estimated to be 306$\pm$11\,$\mu$as, 456$\pm$3\,$\mu$as, 641$\pm$6\,$\mu$as, and
743$\pm$2\,$\mu$as at these four epochs, respectively.
\label{fig:image_2229+695}}
\end{figure}

The closure-only images of source 2229+695 derived from quad-band VLBI
observations provide interesting information to understand these
large variations. We selected images at the highest
frequency band (10.5 GHz) to have the highest angular resolution and at four
epochs between 2019 and 2024 for this purpose (Fig.\,\ref{fig:image_2229+695}).
By performing model
fitting in \emph{difmap}, the structure can be modeled as two or three circular
Gaussian components. Based on these images one could see that the peak
component moves continuously
towards the East with respect to the most Western component, which agrees with
the observed motion of the astrometric position of 2229+695
(Fig.\,\ref{fig:2d_2229+695}). It is therefore likely that the peak component
is a jet component while the weaker, Western component would be the core. This
demonstrates that quad-band VLBI observations have the capability of determining
jet kinematics down to the level of 10\,$\mu$as/yr or better, similar to what
astronomical VLBI observations
can achieve. Among the sources frequently observed by VGOS, others
clearly show such continuous motion in one direction, among which are
0642+449, 1030+415, and 1039+811. One should note that a large position jump 
opposite to the present motion might happen at some point in the future for
these sources if the jet fades and becomes weaker than the core, which
would be fully detectable by the quad-band observations. From an
astrophysical perspective, it would be interesting to investigate why the
jet of these sources has remained brighter than the core for many years and
when it will fade to be weaker than the core. For astrometry and geodesy,
it is crucial to register the images of a source to a feature that
remains stable over time when modeling source structure in
geodetic solutions. In the case of 2229+695, this will mean not using the peak
component for this purpose, but rather the weaker component in the West.

\subsection{Back-and-forth motion along one direction}
\begin{figure}[ht!]
\plotone{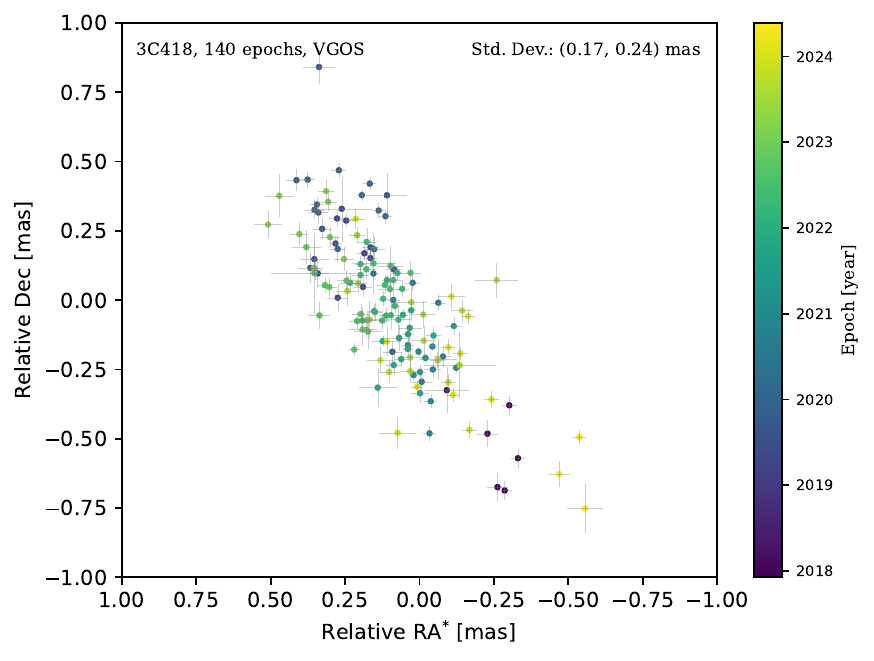}
\gridline{
          \fig{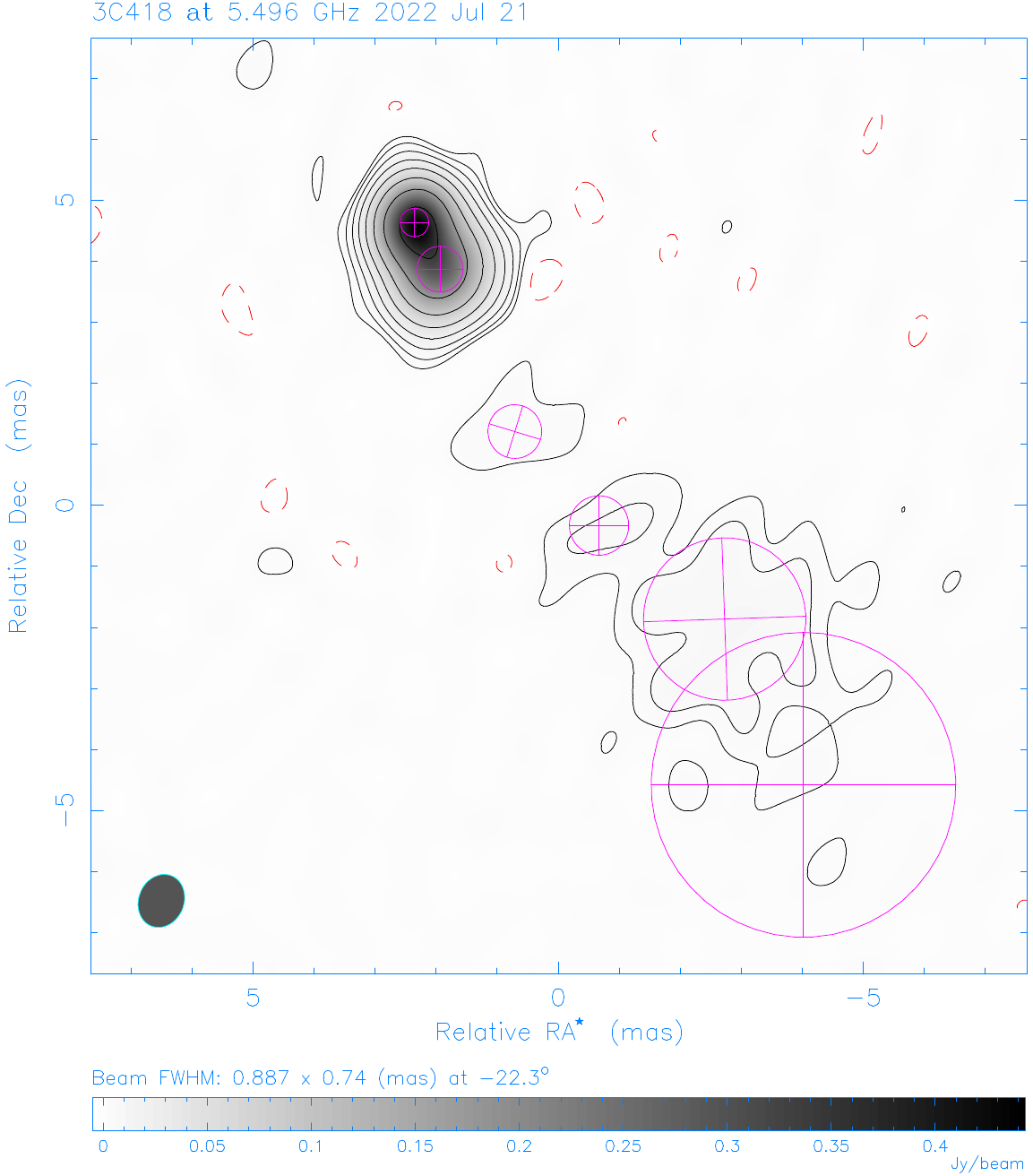}{0.5\textwidth}{}
          \fig{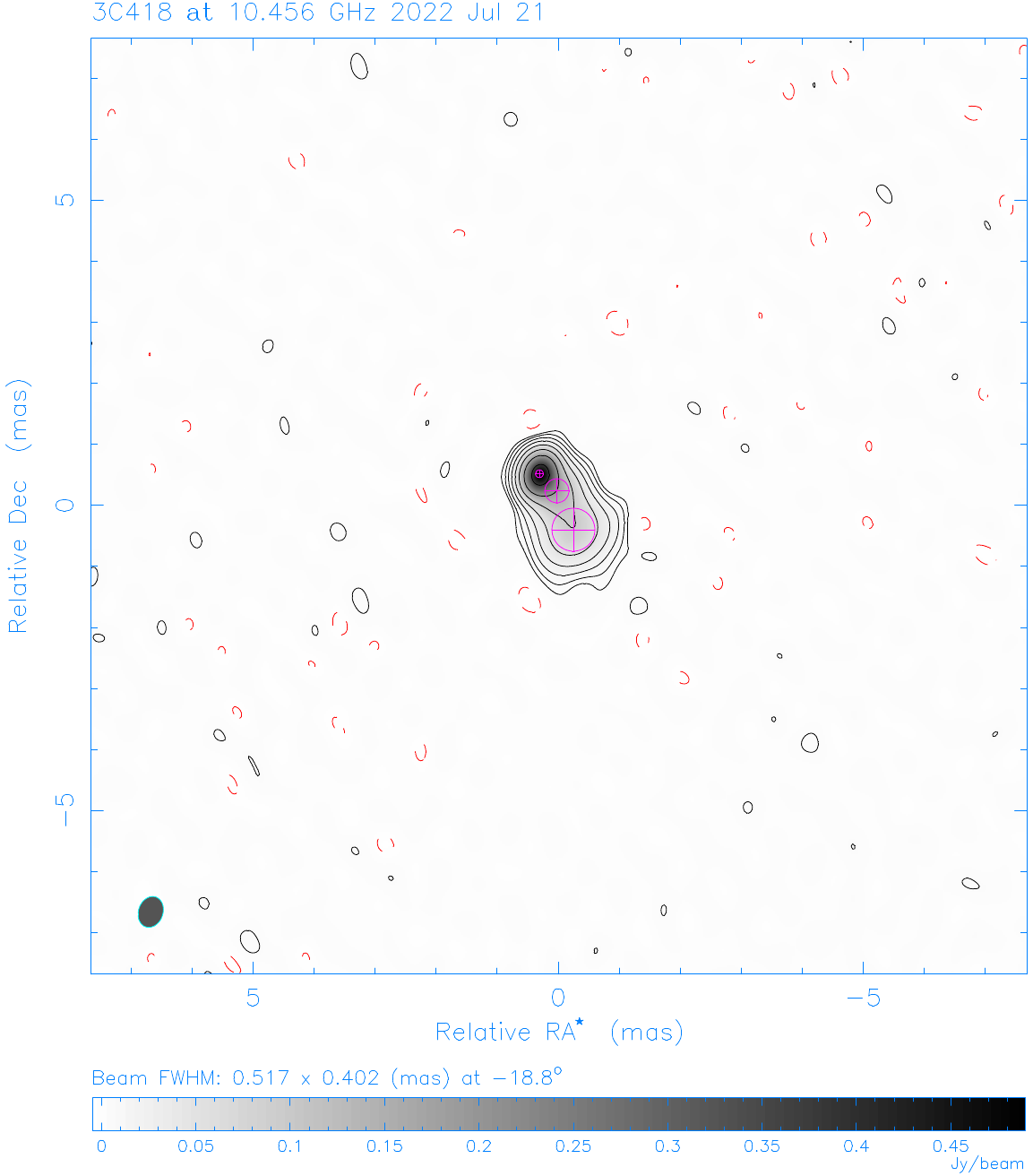}{0.5\textwidth}{}
          }
\caption{Illustration of intrinsic position variations for the source 3C418
along its jet. The top panel shows the variation of its quad-band position on
the sky, while the bottom panel shows the derived closure-only images at
5.5\,GHz (bottom-left) and 10.5\,GHz (bottom-right) from session vr2204.
The ellipse in the bottom-left corner of the image plots shows the beam size.
\label{fig:2d_3C418}}
\end{figure}

Another type of variations common in quad-band source positions is
back-and-forth motion along a direction that is parallel to the
jet direction. To illustrate this, the variation of the
position of 3C418 on the sky is shown in the top panel of
Fig.\,\ref{fig:2d_3C418}
as an example. The position moves from the South-West to the North-East and
then back
to the South-West during 2018 -- 2021. This movement is then repeated during
2021 -- 2024. The position wandering has an amplitude of
$\sim$1.0\,mas in RA$^{*}$ and $\sim$1.2\,mas in Dec. 

The images of 3C418 at 5.5\,GHz and 10.5\,GHz in the bottom panel
of Fig.\,\ref{fig:2d_3C418} show a 10\,mas long elongated jet in the
South-West direction.
The structure at 5.5\,GHz may be modeled as six Gaussian components, with the
inner two components dominating the radio emission, accounting for 84\%
of the total flux density. These two components are almost equally bright and
have an angular distance of 0.82\,mas. This inner structure also
dominates the radio emission at 10.5\,GHz but is resolved into three Gaussian
components at this
frequency band. Due to the lower resolution at 5.5\,GHz, the two innermost
components seen at 10.5\,GHz fall within the extent of the 5.5\,GHz beam
and cannot be distinguished. At the lowest frequency band (3.3\,GHz), all
three components fall within the beam ($\sim$1.6 mas) and are inseparable
(image not shown). In contrast to the visible structure, the structure on scales
smaller than the beam size is largely invisible in the images, and we refer to
it as ``invisible'' structure, again following \citet{2010ivs..conf....8P}. As
pointed out by this author, the response of the visibility phase to such
structure corresponds closely to
that of a point source at the position of the radio brightness
centroid. In this respect, it is worthwhile noting that, for a source like
3C418, the brightness centroid of the invisible structure is shifted towards the
jet further as the observing frequency decreases. As a result, the position
estimates in the four frequency bands would be offset relative to each other if
one could measure them separately. In the same vein, the visibility phases of a
given source refer to different directions on the sky (brightness centroid)
at the four frequency bands. Therefore, the position estimate from the group
delays in quad-band observations is affected by such position offsets across the
four bands. The relationship between these position offsets and the quad-band
position estimates was reported in \citet{2022A&A...663A..83X}.

Back-and-forth motion along one direction such as that observed for 3C418 is
due to the jet kinematics. When jet components emerge from the core, they
first shift the source position towards the jet. Then, once far enough from
the core, these components move from the invisible to the visible region, at
which stage the source position starts to move back to the core. The behavior
will repeat with each new jet component launched. Emergence of new jet
components and jet kinematics are common in AGNs, as
shown, e.g., by the MOJAVE project \citep{2018ApJS..234...12L}. The amplitude
of the position wandering along the jet direction depends on the relative
strength of the core and jet components and thus is different from source to
source. Besides 3C418, this type of motion is found to happen for a
number of sources in our sample (see supplementary material). Other sources
showing such behavior with large amplitudes include 0119+115, 0229+131, and
0133+476.

\section{Evaluating the stability of quad-band source positions}

As we have seen, the session-wise positions for the sources 2229+695 and 3C418
have much larger standard deviations and show systematic patterns compared to
the more noise-like positions observed for the source 1803+784. As these
systematic variations correspond to physical changes in the location of the
radio emission on the sky due to jet kinematics and can cause significant errors
in the determination of source positions, it would be desirable to remove such
variations from the position time series through appropriate modeling to
increase the accuracy of the estimated positions and approach the intrinsic
noise level enabled by quad-band observations such as that found for the source
1803+784. It would not be justified either to consider the position offsets
between the different frequency bands as overall intrinsic errors of the
respective radio catalogs. Instead, we should acknowledge the existence of
source-dependent position differences between the different frequencies caused
by source structure.

\begin{figure}[ht!]
\gridline{
          \fig{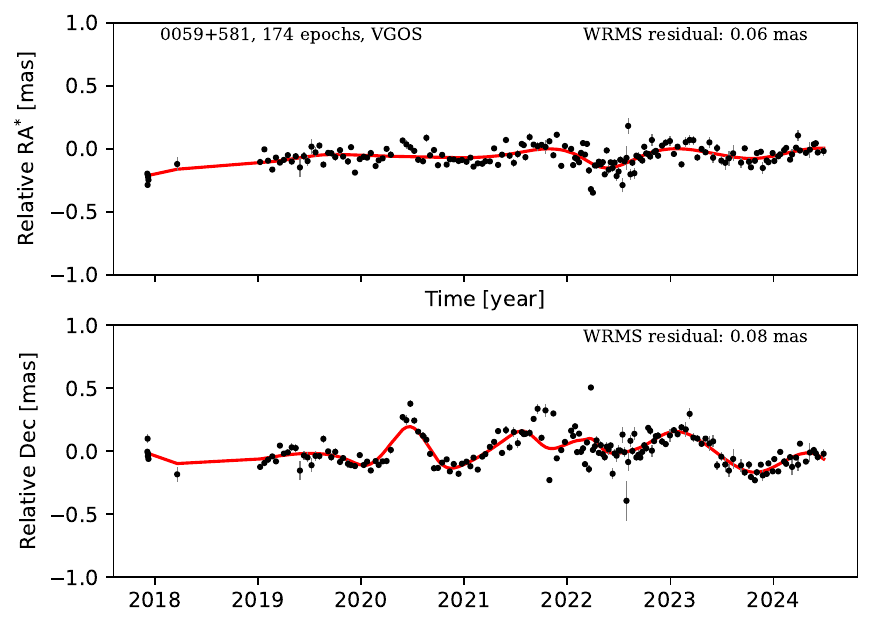}{0.5\textwidth}{}
          }
\caption{Variation of the quad-band coordinates of source 0059+581 over time. A
smoothing model with one order polynomial and a time window of 7 points is
applied to the time series and shown as the red curve. The wrms
value of the residuals with respect to this model is reported in the
upper-right corner of each panel. \label{fig:0059+581}}
\end{figure}

To mitigate the impact of systematic
patterns, we applied to the position time series a smoothing filter with a
first-order polynomial and a time window of 7 points, which corresponds to
$\sim$3 months for the frequently observed sources and a longer period
of time for the sources with sparse observations, given the past rate of
quad-band VLBI sessions. Figure\,\ref{fig:0059+581} shows the
quad-band position time series for source 0059+581, the most frequently
observed source in our dataset to illustrate the impact of the filter. This
source has systematic position variations, mainly in declination, with a
standard deviation of $\sim$120\,$\mu$as. The resulting model is shown as the
red curve. The wrms value of the residuals with respect to the model is
0.06\,mas for RA$^{*}$ and 0.08\,mas for Dec. We note that the smoothing filter
can remove the systematic variations due to source structure to some extent but
one should not expect to do so fully. This is because the beam size and thus the
extent of the in-beam structure depends also on the size of the observing
network and can change dramatically from session to session, which could then
lead to sudden jumps in the observed positions. While these may be recognized as
network effects, they are actually source-structure driven.

\begin{figure}[ht!]
\gridline{
          \fig{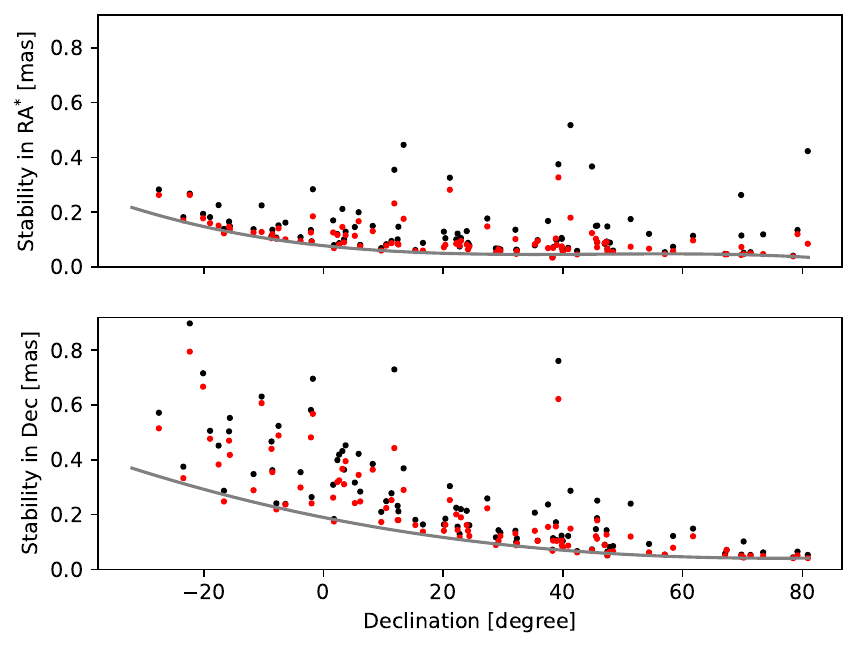}{0.5\textwidth}{}
          }
\caption{Stability of the quad-band source positions as a function of
declination for RA$^{*}$ (upper panel) and for Dec (lower panel). The standard
deviations of the
coordinate time series with respect to the weighted means are shown as black
dots whereas the wrms values of the coordinate residuals with respect to the
smoothing
models are shown as red dots. The lower edge of the distribution of the red dots
characterizes the precision that one can reach at most in the determination of
 source positions and is shown as a grey curve in each plot. The instability
above the gray curve is mainly caused by source structure, at least for the
sources with declination above 20$\degree$, whereas the gray curve itself
indicates the level of the error contributions due to other effects, such as
tropospheric path delays and limitation on the sensitivity of the VGOS network
to position coordinates.
\label{fig:std_dec}}
\end{figure}

By using the smoothing filter, we explored the position time series for 96
sources observed in at least 30 quad-band VLBI sessions. The standard
deviations of the original time series for these 96 sources have a median value 
of 0.11\,mas for RA$^{*}$ and 0.19\,mas for Dec, and the wrms values of the 
residuals with respect to the derived models have a median of 0.08\,mas for
RA$^{*}$ and 0.15\,mas for Dec. Figure\,\ref{fig:std_dec} reports these 
values as a function of declination. The stability in RA$^{*}$ increases
significantly after removing the systematic variations.
This also happens in declination for sources with declination
above 10$\degree$. It is very unlikely that factors like tropospheric
effects are able to cause systematic variations in source positions with a time
window of three months. This suggests that, as demonstrated by our
investigations on 2229+695
and 3C418, source structure may be the major factor causing position
variations for the sources in the North.

\begin{figure}[ht!]
\gridline{
          \fig{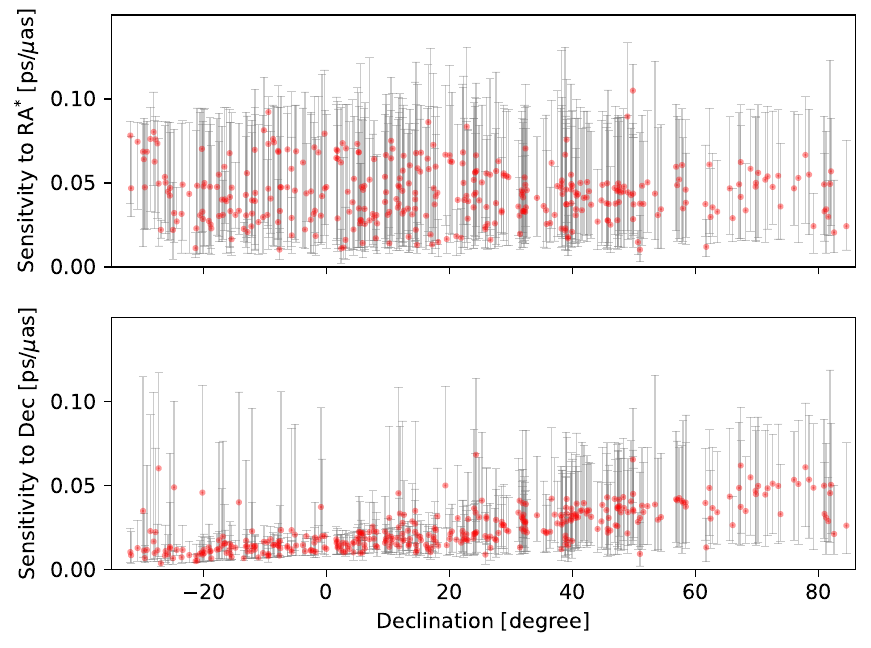}{0.5\textwidth}{}
          }
\caption{Sensitivity of the existing quad-band VLBI
observations to source coordinates as a function of declination. For each
source, the median absolute value of the partial derivatives of the group
delays with respect to each of the two coordinates, RA$^{*}$ and Dec, is
calculated based on the actual observations of the 177 sessions and is shown by
a red dot. The lower and upper bars represent the first and third quantiles of
the absolute values of the partial derivatives, respectively.
\label{fig:sensitivity}}
\end{figure}

Besides the effects induced by systematic position variations,
Fig.\,\ref{fig:std_dec} clearly shows the strong dependence of the
position stability on declination. The higher instability in the Southern
hemisphere results from the asymmetry of the current VGOS network between
the Northern and Southern hemispheres. Presumably, there are two reasons for
the instability of sources
with declinations below 20$\degree$: (1) the geometry of the observations
is less sensitive to the source positions and (2) the
atmospheric path delays become more prominent since the observations are
conducted at lower
elevation angles. Due to Earth
rotation, VLBI observations have nearly uniform sensitivity to right
ascension. However, for declination, the East-West
baselines provide very little sensitivity for sources with low declinations.
Although the North-South baselines do provide sensitivity to
declination, the current VGOS network does not have a
sufficient number, if any, of long baselines in the North-South direction to
match the sensitivity on right ascension, hence the asymmetry found between the
two coordinates.
The sensitivity of VLBI observations to source coordinates can be
quantitatively assessed by the partial derivatives of the group delays with
respect to right ascension and declination. To evaluate the
impact of the two factors (network geometry and atmospheric effects) on the
instability of sources with declinations
below 20$\degree$, we plotted the median,
together with the first and third quantiles, of the absolute values of the
partial derivatives of the group delays with respect to the source coordinates
for the 177 quad-band experiments (see Fig.\,\ref{fig:sensitivity}).
The median sensitivity to RA$^{*}$ is on average
about 0.05\,ps/$\mu$as whatever the declination. While there is a large
fluctuation in this median sensitivity from source to source, the third
quantile is uniform, at the level of 0.1\,ps/$\mu$as.
However, the sensitivity to Dec decreases significantly when declination goes
down, with median values ranging from 0.05\,ps/$\mu$as at the
North Pole to 0.003\,ps/$\mu$as for declinations lower than $-$20$\degree$.
This suggests that the observation sensitivity is the dominant factor for
the high instability in Dec but not for that in RA$^{*}$. The level of the
instability in RA$\*$, about 0.2\,mas, can then be used as a measure of the
impact of systematic errors in delay observations, such as atmospheric path
delays, on source coordinates. In the same way, the level of the instability in
Dec, about 0.5\,mas, may be used to quantify the impact of the observation
sensitivity on the declination coordinate, provided that atmospheric path
delays cause errors in the two coordinates at the same level. Expanding
the VGOS network towards the equator and into the South is an urgent need. We
also note that there is no dependence of the position stability on right
ascension.

The lower envelope of the distribution of
the wrms residuals in right ascension and declination as a function of
declination
(Fig.\,\ref{fig:std_dec}) characterizes the stability
of the source coordinates in a global solution estimating session-wise
positions. It is about 40\,$\mu$as for both RA$^{*}$ and Dec at the highest
declinations but increases to $>$ 0.32\,mas in RA$^{*}$ and $>$ 0.47\,mas in
Dec at
declinations below $-$45$\degree$. We refer to this lower envelope as the
error floor in quad-band source positions. Instead of using a single
value for all radio
sources to inflate the coordinate uncertainties, we use this
declination-dependent error floor
to determine the coordinate uncertainty for each source. This error
floor is tabulated with a 15$\degree$ interval in declination and reported in
Table \ref{t:stability}. As discussed, the
observation sensitivity to source positions and atmospheric path delays
cause significant declination-dependent instability in the quad-band positions.
The identified noise floor has taken these two factors into account. However, it
does not take into account
the systematic variations due to source structure (since the noise floor is the
lower envelope of the stability distribution). The rationale for doing so is
based on the facts that: (1) systematic variations due to source structure
correspond to changes in the location of radio emission and can be treated
as real signals in the position catalogs at various frequency bands, rather
than as errors, and (2) magnitude of such systematic variations is strongly
source-dependent, which poses a challenge in inflating the position uncertainty
for each source to fairly take them into account.

%

\begin{deluxetable*}{lcccccccccc}
\tablenum{1}
\tablecaption{Error floor in quad-band source positions as
a 
function of declination (units: milli-arcsecond) \label{t:stability}}
\tablewidth{0pt}
\tablehead{
&\colhead{$-45^{\circ}$}&\colhead{$-30^{\circ}$}&\colhead{$-15^{\circ}$}&\colhead{$0^{\circ}$}
&\colhead{$15^{\circ}$}&\colhead{$30^{\circ}$}&\colhead{$45^{\circ}$}&\colhead{$60^{\circ}$}
&\colhead{$75^{\circ}$}&\colhead{$90^{\circ}$}
}
\startdata
RA$^{*}$ & 0.32 & 0.20 & 0.13 & 0.08 & 0.05 & 0.05 & 0.05 & 0.05 & 0.04 & 0.03\\
Dec      & 0.47 & 0.36 & 0.26 & 0.19 & 0.13 & 0.09 & 0.06 & 0.05 & 0.04 & 0.04 \\
\enddata
\end{deluxetable*}

\section{Differences between quad-band positions and S/X
positions} \label{sec:difference}

In geodetic analysis, least-squares adjustment is used to determine the
estimates of the parameters and their uncertainties based on the assumptions
that the un-modeled errors are purely thermal noise and the covariance matrix of
the observations consists of non-zero values only in the main diagonal -- the
observations are not correlated. In general, these two assumptions are not
fulfilled in
VLBI data analysis, which leads to systematic errors in the parameter estimates
and an underestimation of their uncertainties. In this study, we used the error
floor reported in Table\,\ref{t:stability} to inflate the
uncertainties of the quad-band source positions determined from the global 
solution. Depending on the declination of the source, the
corresponding error floor is interpolated. It is scaled down with the square
root of the number of observing sessions and added in quadrature to the
original uncertainty for both RA$^{*}$ and Dec from the global solution. The
scaling with the number of sessions is driven by the fact that the error floor,
determined from position time series, is valid for a single
quad-band session. The number of sessions for a source is counted based on the
sessions that have more than 500 good observations for that source. Sessions
with fewer observations are considered also but only counted as
fractions of a session. While this threshold of 500 might need further
justification, it is not critical as long as this number remains between 100
and 1000.

\begin{figure}[ht!]
\gridline{
          \fig{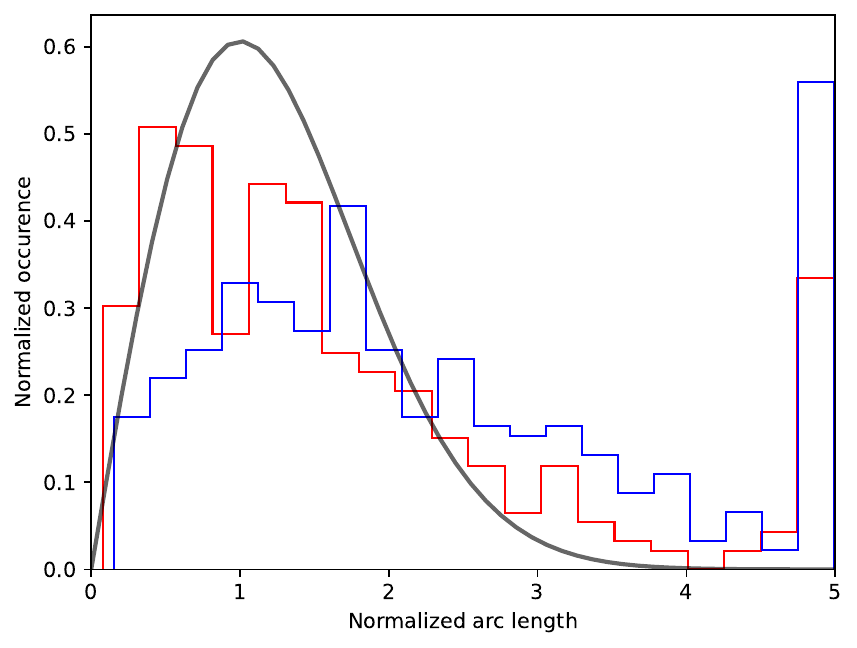}{0.5\textwidth}{}
          }
\caption{Distribution of normalized arc lengths between the quad-band and S/X 
positions. The blue bars correspond to the normalized arc lengths based on 
the original uncertainties of the quad-band positions whereas the red bars
correspond to those derived with inflated uncertainties. The normalized arc
lengths exceeding 5.0 are counted in the last bin. The Rayleigh distribution is
shown as the gray curve. \label{fig:arc_length}}
\end{figure}

We calculated the arc length, denoted by $\rho$, between the
quad-band position and the S/X position of each source as
\begin{equation}
\label{rho}
\rho = \sqrt{\Delta{\text{RA}^{*}}^{2}+\Delta{\text{Dec}}^{2}},
\end{equation}
where $\Delta\text{RA}^{*}$ and $\Delta\text{Dec}$ are the differences in 
RA$^{*}$ and Dec between the quad-band and S/X positions. The normalized arc 
length, $X_{\rho}$, is then defined as
\begin{equation}
\label{X_rho}
X_{\rho} = \rho/\sigma_{\rho},
\end{equation}
where $\sigma_{\rho}$ is the uncertainty of $\rho$. This uncertainty is
calculated by projecting the uncertainties of RA$^{*}$ and Dec to the direction
of the vector joining the two positions, taking their
correlation into account, and this is
done for both the quad-band and S/X position uncertainties \citep[see,
e.g.,
][]{2019MNRAS.482.3023P}. The distribution of the
normalized arc lengths is shown in Fig.\,\ref{fig:arc_length} when considering
either the original uncertainties or the inflated
uncertainties of the quad-band positions. It is obvious that the inflated
uncertainties provide an arc length distribution that is closer to the ideal
Rayleigh distribution. However, one should not expect it to exactly follow
the Rayleigh distribution unless source
structure is modeled. Inflating the
position uncertainties so that the arc length distribution matches the Rayleigh
distribution, as done by \citet{2024AJ....168...76P}, would
not be enough to fully characterize all the factors that
affect source position estimates and would just lead to hiding the differences
between the positions at the different bands. With respect to the S/X
positions, 15\% of the sources have quad-band
positions that are offset by more than 3$\sigma$, with median and mean
differences of
0.58\,mas and 0.82\,mas, respectively. These radio sources are reported in
Table\,\ref{arc_length}. Considering the sources with declinations above
20$\degree$, 22\% have significant position offsets. In general, these
sources have prominent jets, as is
the case of 0723$-$008 discussed in the next paragraph.

\begin{deluxetable*}{lcrDDDDDD}
\tablenum{2}
\tablecaption{List of the 59 sources for which the quad-band and S/X band
positions differ by more than 3$\sigma$.
\label{arc_length} }
\tablewidth{0pt}
\tablehead{
\colhead{Source} & \colhead{RA} & \colhead{Dec} & \multicolumn2c{$\Delta 
\textrm{RA}^{*}$} & \multicolumn2c{$\Delta \textrm{Dec}$} & 
\multicolumn2c{${\sigma}_{\text{RA}^{*}}$} & 
\multicolumn2c{${\sigma}_{\textrm{Dec}}$} &  \multicolumn2c{Arc length}  & 
\multicolumn2c{Normalized} \\
\colhead{}& \colhead{[deg.]} & \colhead{[deg.]} & \multicolumn2c{[mas]} & 
\multicolumn2c{[mas]} & \multicolumn2c{[mas]} & \multicolumn2c{[mas]} & 
\multicolumn2c{[mas]} & \multicolumn2c{arc length}
}
\decimalcolnumbers
\startdata
0003$-$066 &    1.5578870 &   $-$6.3931488 &   0.04 &  -0.29 &   0.05 &   0.09 &
0.29 &  3.5 \\
0014+813 &    4.2853124 &   81.5855931 &   0.15 &  -1.19 &   0.06 &   0.06 &
1.20 & 14.3 \\
0016+731 &    4.9407763 &   73.4583382 &  -0.28 &   0.01 &   0.03 &   0.03 &
0.28 &  6.3 \\
0119+115 &   20.4233128 &   11.8306705 &   0.47 &   0.67 &   0.03 &   0.05 &
0.82 & 13.2 \\
0146+056 &   27.3432124 &    5.9315466 &   1.25 &  -0.83 &   0.12 &   0.29 &
1.50 &  5.5 \\
0202+149 &   31.2100579 &   15.2364008 &   0.17 &  -0.76 &   0.06 &   0.15 &
0.78 &  6.1 \\
0212+735 &   34.3783896 &   73.8257281 &   0.68 &  -0.69 &   0.06 &   0.06 &
0.97 & 10.8 \\
0229+131 &   37.9412255 &   13.3818657 &   0.88 &   0.09 &   0.04 &   0.07 &
0.88 & 10.0 \\
0319+121 &   50.4712645 &   12.3538763 &  -0.30 &   0.81 &   0.10 &   0.20 &
0.87 &  4.5 \\
NRAO140  &   54.1254481 &   32.3081507 &  -0.74 &   0.39 &   0.08 &   0.12 &
0.84 &  5.7 \\
NRAO150  &   59.8739470 &   50.9639337 &   0.12 &  -0.45 &   0.05 &   0.05 &
0.46 &  7.0 \\
0434$-$188 &   69.2561778 &  $-$18.7468366 &  -1.07 &   1.93 &   0.17 &   0.36 &
2.21 &  7.2 \\
0454+844 &   77.1765149 &   84.5345957 &   0.09 &   0.33 &   0.05 &   0.05 &
0.34 &  4.6 \\
0620+389 &   96.0792553 &   38.9468712 &  -0.31 &   0.27 &   0.08 &   0.10 &
0.41 &  3.0 \\
0642+449 &  101.6334415 &   44.8546083 &  -0.30 &  -0.12 &   0.03 &   0.03 &
0.32 &  6.8 \\
0650+371 &  103.4928450 &   37.0946129 &  -0.50 &  -0.48 &   0.09 &   0.11 &
0.70 &  4.7 \\
0723$-$008 &  111.4609989 &   $-$0.9157053 &  -2.55 &   3.99 &   0.16 &   0.28 &
4.73 & 16.5 \\
0738+313 &  115.2945971 &   31.2000644 &   0.10 &   2.58 &   0.08 &   0.11 &
2.58 & 21.3 \\
0833+585 &  129.3433734 &   58.4171791 &  -0.42 &  -0.07 &   0.09 &   0.08 &
0.42 &  3.5 \\
0850+581 &  133.6749847 &   57.9583166 &  -0.63 &   0.73 &   0.08 &   0.09 &
0.96 &  7.9 \\
0912+029 &  138.6579727 &    2.7664572 &   0.24 &  -0.56 &   0.10 &   0.24 &
0.61 &  3.1 \\
0917+449 &  140.2435771 &   44.6983291 &   0.12 &  -0.27 &   0.05 &   0.06 &
0.30 &  3.6 \\
0919$-$260 &  140.3723079 &  $-$26.3120522 &   0.47 &  -1.53 &   0.21 &   0.64 &
1.60 &  3.0 \\
0917+624 &  140.4009630 &   62.2644945 &   0.27 &  -0.07 &   0.06 &   0.06 &
0.28 &  3.2 \\
4C39.25  &  141.7625584 &   39.0391255 &   0.86 &  -0.10 &   0.06 &   0.08 &
0.87 &  7.9 \\
OK290    &  149.2078142 &   25.2544584 &   0.49 &   0.51 &   0.08 &   0.13 &
0.70 &  4.8 \\
1030+415 &  158.2654494 &   41.2683981 &  -0.31 &   0.22 &   0.03 &   0.04 &
0.38 &  7.7 \\
1039+811 &  161.0960945 &   80.9109564 &   0.25 &   0.00 &   0.03 &   0.03 &
0.25 &  5.5 \\
1059+282 &  165.5595354 &   27.9524137 &   0.45 &  -0.30 &   0.05 &   0.06 &
0.54 &  7.1 \\
1213$-$172 &  183.9447992 &  $-$17.5292787 &   0.48 &  -0.08 &   0.06 &   0.10 &
0.48 &  3.9 \\
1307+121 &  197.3913853 &   11.9068204 &   0.37 &   0.51 &   0.10 &   0.19 &
0.63 &  3.2 \\
1308+326 &  197.6194328 &   32.3454951 &   0.38 &  -0.42 &   0.06 &   0.10 &
0.57 &  5.0 \\
1319$-$093 &  200.6538025 &   $-$9.6271663 &  -0.34 &   1.63 &   0.15 &   0.35 &
1.67 &  5.8 \\
1354+195 &  209.2684863 &   19.3187143 &   0.62 &  -0.76 &   0.10 &   0.18 &
0.99 &  5.1 \\
OQ208    &  211.7516433 &   28.4540803 &  -0.48 &  -1.05 &   0.06 &   0.11 &
1.16 &  9.6 \\
1418+546 &  214.9441559 &   54.3874408 &   0.14 &  -0.09 &   0.03 &   0.03 &
0.16 &  3.6 \\
1448+762 &  222.1199134 &   76.0198882 &   0.48 &   0.14 &   0.06 &   0.06 &
0.50 &  5.6 \\
1508$-$055 &  227.7232976 &   $-$5.7187274 &   0.09 &  -0.99 &   0.13 &   0.40 &
0.99 &  3.1 \\
1510$-$089 &  228.2105538 &   $-$9.0999524 &  -0.21 &   1.00 &   0.16 &   0.41 &
1.03 &  3.0 \\
1547+507 &  237.3227862 &   50.6349409 &   1.21 &  -0.82 &   0.37 &   0.20 &
1.46 &  3.1 \\
1548+056 &  237.6469552 &    5.4529021 &   0.11 &  -0.91 &   0.11 &   0.23 &
0.92 &  4.4 \\
1557+032 &  239.8790525 &    3.0800714 &  -0.21 &   0.44 &   0.07 &   0.15 &
0.49 &  3.5 \\
1637+574 &  249.5560680 &   57.3399942 &   0.14 &   0.28 &   0.06 &   0.07 &
0.31 &  3.3 \\
1642+690 &  250.5327024 &   68.9443768 &   0.28 &  -0.06 &   0.05 &   0.05 &
0.29 &  3.8 \\
\enddata
\tablecomments{Column (4) reports the angular difference in right ascension,
i.e., scaled by $\cos$(Dec). The same scaling was applied to its uncertainty
in column (6).}
\end{deluxetable*}

\begin{deluxetable*}{lcrDDDDDD}
\tablenum{2}
\tablecaption{Continued\label{arc_length} }
\tablewidth{0pt}
\tablehead{
\colhead{Source} & \colhead{RA} & \colhead{Dec} & \multicolumn2c{$\Delta 
\textrm{RA}^{*}$} & \multicolumn2c{$\Delta \textrm{Dec}$} & 
\multicolumn2c{${\sigma}_{\text{RA}^{*}}$} & 
\multicolumn2c{${\sigma}_{\textrm{Dec}}$} &  \multicolumn2c{Arc length}  & 
\multicolumn2c{Normalized} \\
\colhead{}& \colhead{[deg.]} & \colhead{[deg.]} & \multicolumn2c{[mas]} & 
\multicolumn2c{[mas]} & \multicolumn2c{[mas]} & \multicolumn2c{[mas]} & 
\multicolumn2c{[mas]} & \multicolumn2c{arc length}
}
\decimalcolnumbers
\startdata
3C345    &  250.7450418 &   39.8102762 &   0.76 &   0.21 &   0.08 &   0.09 &
0.79 &  6.3 \\
1745+624 &  266.5584753 &   62.4485383 &  -0.42 &  -0.38 &   0.06 &   0.06 &
0.57 &  7.1 \\
1751+288 &  268.4269735 &   28.8013720 &   0.06 &   0.16 &   0.03 &   0.03 &
0.17 &  3.7 \\
1800+440 &  270.3846450 &   44.0727500 &  -0.12 &  -0.27 &   0.05 &   0.05 &
0.30 &  4.4 \\
1846+322 &  282.0920357 &   32.3173899 &   0.05 &  -0.14 &   0.03 &   0.04 &
0.15 &  3.1 \\
1947+079 &  297.5230828 &    8.1205513 &  -0.77 &   3.62 &   0.14 &   0.27 &
3.70 & 14.6 \\
1954+513 &  298.9280763 &   51.5301517 &   0.44 &  -0.16 &   0.07 &   0.07 &
0.47 &  4.6 \\
2000+472 &  300.5434095 &   47.4246593 &   0.27 &  -0.14 &   0.03 &   0.03 &
0.30 &  6.4 \\
2007+777 &  301.3791609 &   77.8786799 &   0.36 &   0.02 &   0.04 &   0.05 &
0.36 &  5.6 \\
2113+293 &  318.8725560 &   29.5606575 &  -0.12 &  -0.09 &   0.03 &   0.03 &
0.15 &  3.2 \\
2201+315 &  330.8123990 &   31.7606304 &  -0.30 &  -0.54 &   0.07 &   0.11 &
0.61 &  4.9 \\
2214+350 &  334.0833746 &   35.3039389 &  -0.03 &   0.22 &   0.05 &   0.06 &
0.23 &  3.2 \\
2229+695 &  337.6519573 &   69.7744658 &   0.08 &  -0.12 &   0.03 &   0.03 &
0.14 &  3.2 \\
2234+282 &  339.0936284 &   28.4826146 &  -0.41 &  -0.60 &   0.07 &   0.12 &
0.73 &  5.4 \\
2320$-$035 &  350.8831406 &   $-$3.2847285 &  -0.26 &   1.11 &   0.16 &   0.42 &
1.14 &  3.4 \\
\enddata
\end{deluxetable*}

\begin{figure}[ht!]
\gridline{
          \fig{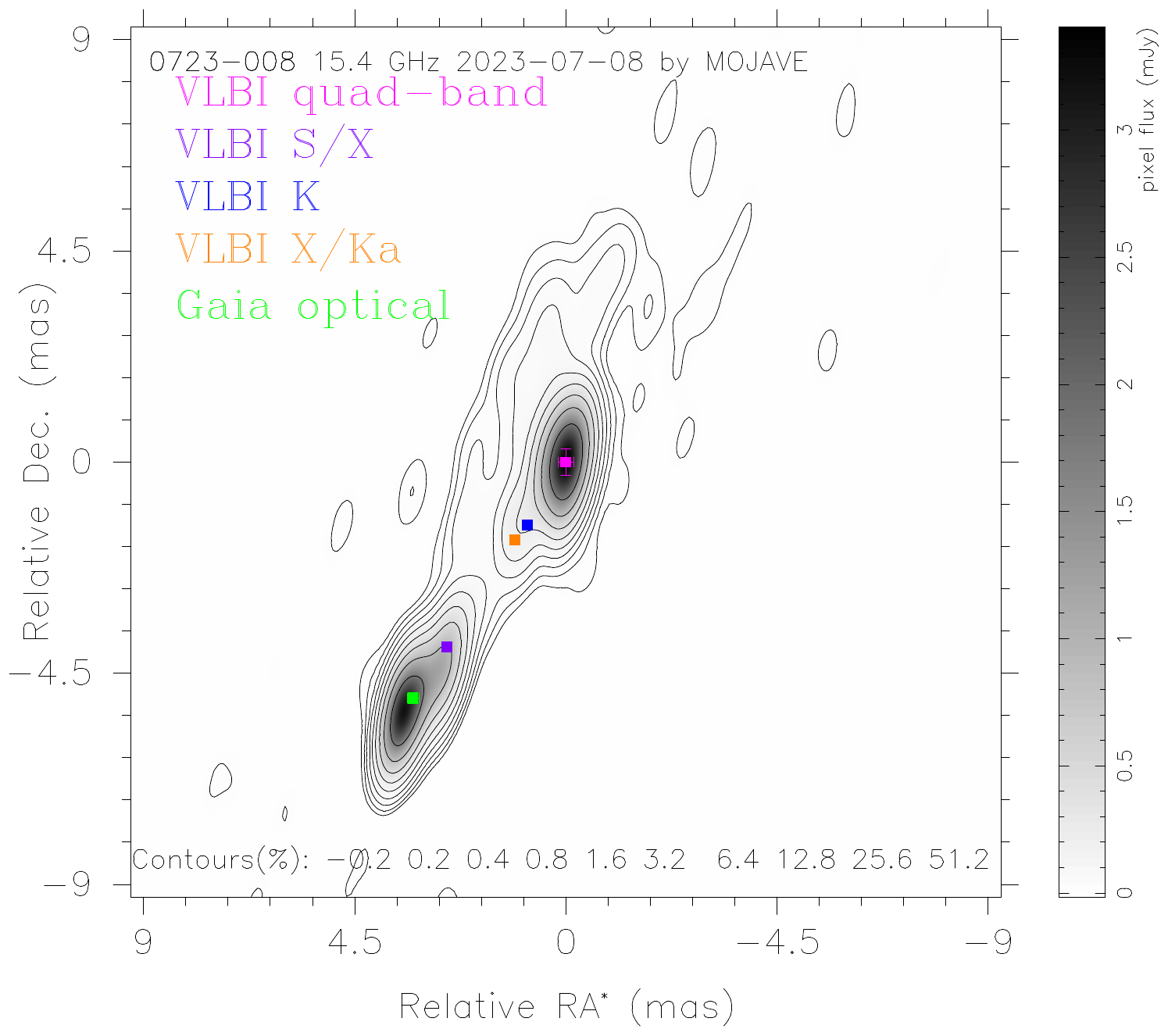}{0.5\textwidth}{}
          }
\caption{Relative positions of source 0723$-$008 at
optical and radio wavelengths with a MOJAVE radio image at 15 GHz in
the background. The MOJAVE image selected is from 8 July 2023 and is the
closest in time to session vr2302 (carried out on 31 May 2023), the only
quad-band session where source 0723$-$008 was observed.
\label{fig:0723-008} }
\end{figure}

The largest difference between the quad-band and S/X positions happens for the
source 0723$-$008, which has an arc length of 4.73$\pm$0.29 mas. To explain
this difference, the absolute positions from radio and optical observations are
plotted in Fig.\,\ref{fig:0723-008}. On the background, a 15\,GHz radio image
from MOJAVE \citep{2018ApJS..234...12L} is shown (since the current quad-band
VLBI observations do not permit obtaining a good-quality image for this source
from our current analysis). The image is for an epoch that is close to session
vr2302 (that was carried out one month earlier), the only VGOS session where
0723$-$008 was observed so far and from which its quad-band position was
derived. Based on long-term monitoring from the MOJAVE project, the
South-Eastern component was identified as the core, while the brighter
component in the North-West corresponds to a jet component. This morphology
contrasts with that seen at earlier epochs, either in MOJAVE images or at S/X
band \citep{1997ApJS..111...95F}, where the North-Western component was much
weaker or even not detected. As absolute
positions are lost in the imaging process, we chose to place the quad-band
position at the location of the brightest component in the image (i.e., the
North-Western jet component). The absolute VLBI positions at S/X, K and X/Ka
bands \citep{2020A&A...644A.159C} along with the Gaia optical position
\citep{2022A&A...667A.148G} are pinpointed on the image accordingly based
on their offsets relative to the quad-band position. With this alignment (i.e.,
the quad-band position corresponding to the location of the
brightest component), the optical position is found to coincide with the core
component while the radio positions at S/X, K and X/Ka bands fall within the
jet (Fig.\,\ref{fig:0723-008}). It is striking that the offset between the
quad-band position and the optical position, with a magnitude of as large as
4.7 mas, agrees with the angular distance between the two prominent
radio components. By examining earlier MOJAVE images, it appears that the
North-Western jet component was ejected from the core at around 2010 and has
moved to its current location since then. It is interesting that the S/X
position, which has the longest observing history (from 1981 to 2018), is closer
to the core, while the K and X/Ka band positions which rely on more recent
observations (2016--2017 for K band and 2005--2018 for X/Ka band) are shifted to
the North-West, which is consistent with a scenario where the brighter
North-Western jet component, after its ejection, has drawn the source position
to the North-West in the past decade or so. Considering all elements, there are
strong indications that the quad-band position corresponds to the location of
the North-Western jet component (as was originally assumed) and that the Gaia
position corresponds to the location of the core.

\begin{figure}[ht!]
\gridline{
          \fig{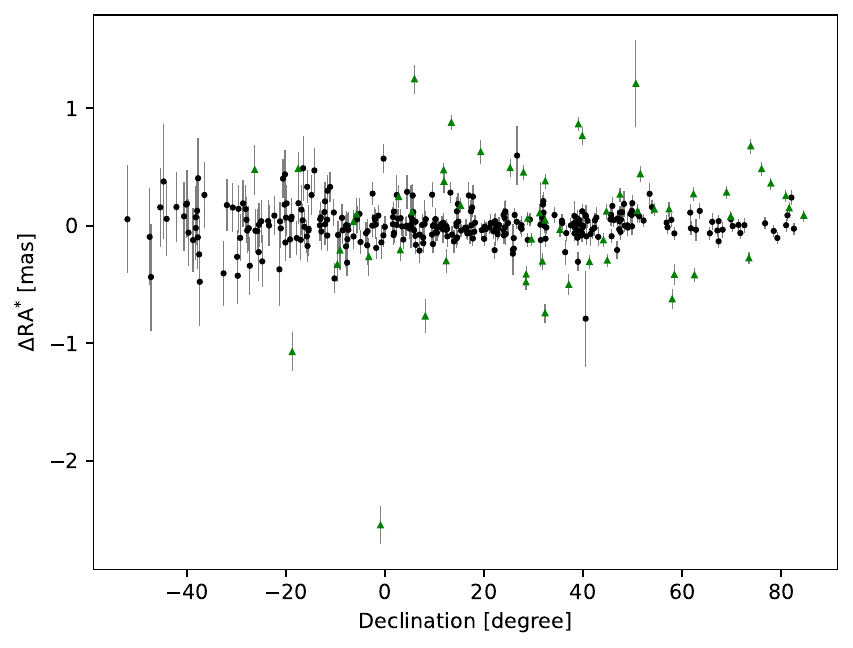}{0.5\textwidth}{}
          \fig{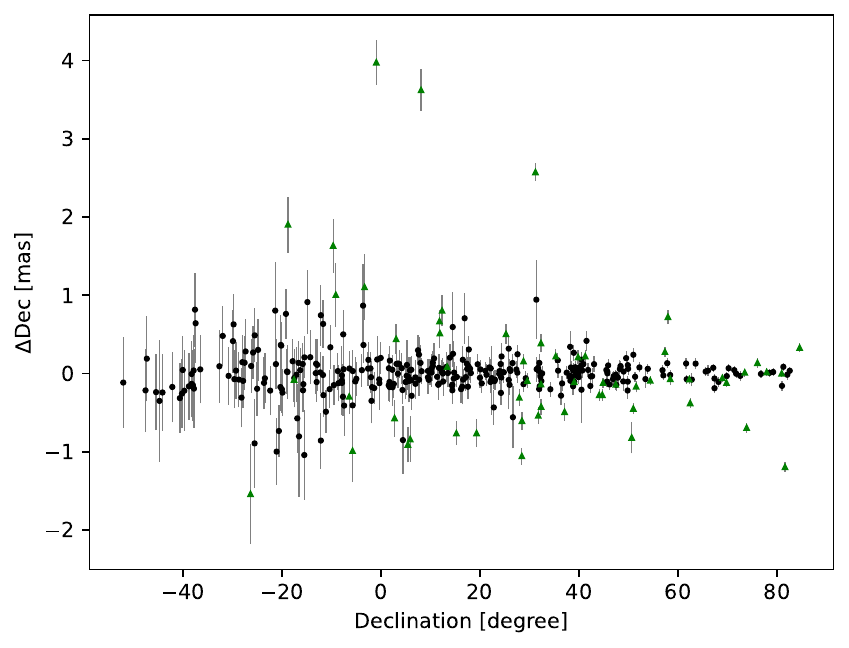}{0.5\textwidth}{}
          }
\caption{Source position offsets between quad-band and S/X band as a function
of declination. The error bars are calculated by adding in quadrature the 
uncertainties of the S/X positions in the ICRF3 to the inflated uncertainties
of the quad-band positions. The green triangles highlight the position offsets 
that are significant at the 3$\sigma$ level. \label{fig:relative}}
\end{figure}

The relative positions between quad-band and S/X band are shown in
Fig.\,\ref{fig:relative} as a function of declination. The median and mean arc
lengths for the 377 sources are
0.175\,mas and  0.310\,mas, respectively. There are 22 sources, about 6\% of
the catalog, with arc lengths
larger than 0.8\,mas and significant at the 3$\sigma$ level. Note that these 6\%
do not include sources like 2229+695 or 3C418 for which the averaged positions
show small differences even though there are large position offsets detected at
some epochs. It is worth noting that there were nearly no such large
position offsets in the results from \citet{2024AJ....168...76P} (see
Figs. 8 and 9 therein). This could be due to the different strategy applied
in the session-wise analysis (see Sect. \ref{sec:session-wised}). Our results
demonstrate the necessity of establishing a specific position catalog
for the quad-band VLBI observations.

One should note that the differences in source positions between various
frequencies most likely stem from source structure and its evolution over
time and frequency. Other conditions related to the coverage and geometry of
the observations like time spans and antenna networks, can also cause position
differences (whether they originate from source structure or not), especially
considering that the quad-band dataset and antenna network are still limited in
this early stage. One cannot draw a final conclusion that the observing
frequency is the only or main factor for these differences.

\section{Quad-band source position catalog} \label{sec:highlight}

\begin{figure}[ht!]
\gridline{
          \fig{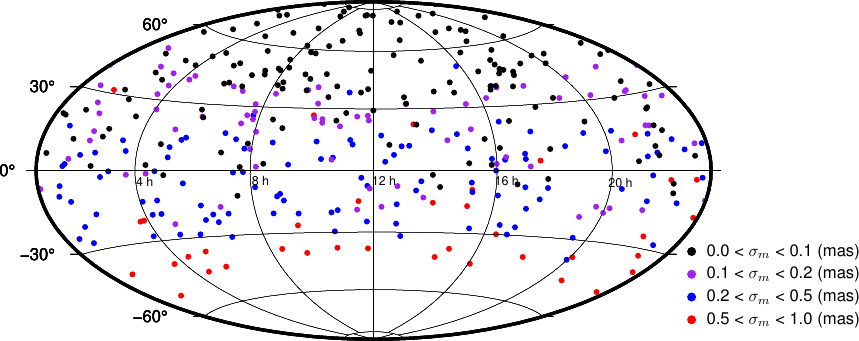}{0.7\textwidth}{}
          }
\caption{Sky distribution of 377 radio sources in the quad-band position
catalog. The color coding depicts the length of the semi-major axis of the
position error ellipse in four predefined ranges, as printed in the
bottom-right.
\label{fig:crf} }
\end{figure}

The final global solution was done by selecting 46 radio sources to
define the CRF datum based on two criteria: (1) observed in at least five
sessions and (2) with a standard deviation (calculated from the coordinate time
series) that is smaller than 0.1\,mas for RA$^{*}$ and 0.2\,mas for Dec. As
stated above, we parameterized
the positions of seven radio sources that show systematic
variations in their position time series as session-wise parameters, instead
of global parameters over the entire dataset. The position uncertainties 
were inflated as discussed in the previous section. 
When using the ICRF3 S/X coordinates as a priori, 99 sources were found
to have an RA estimate that is significant at the 3$\sigma$ level (mean value
of 0.30\,mas), while 69 sources are in this situation for the Dec estimates
(mean value of 0.25 mas). Overall, 125 sources show significant position
estimates with respect to the S/X positions. This number is twice that found
when comparing the quad-band and S/X catalogs (see
Sect.\,\ref{sec:difference}), which suggests that quad-band VLBI observations
provide more accurate positions than the legacy VLBI
observations at S/X for some sources. Source
1803+784 is one of such cases, as demonstrated by Fig.\,\ref{fig:2d_1803+784}
where an offset of about 0.04\,mas is detected. This justifies
using the source positions from quad-band VLBI observations for the
data analysis of, for instance, quad-band \emph{intensive} observations, where
source positions can be one of the error sources because they are not estimated
in the analysis \citep[e.g.,
][]{2023JGeod..97...97K}. The source positions from the
global solution with inflated uncertainties are reported in Table
\ref{Tab:VGOS_coordinates}, and the sky distribution is shown in
Fig.\,\ref{fig:crf}. This catalog includes modeling for Galactic aberration
\citep[e.g.,][]{2012A&A...544A.135X,2020A&A...644A.159C,2021A&A...649A...9G} and
has a time epoch of 2015.0, in agreement with the ICRF3.
The median uncertainty of the quad-band positions is
0.06\,mas for RA$^{*}$ and 0.10\,mas for Dec. Despite being inflated, the
uncertainties reported here could still be
underestimated because they do not incorporate systematic variations due to
source structure.

\section{Discussion on the impact of source structure}

The investigation of source position variations in Sect.\,\ref{sec:variations}
illustrates the fact that for core-dominated sources it is not the visible
structure (i.e., the structure on large scales) that affects significantly the
source positions determined from absolute astrometry but the invisible
structure (i.e, within the beam size). Given the angular resolutions at the
four VGOS frequency bands (0.4 -- 2.0\,mas)
and the fact that all AGNs are resolved on these scales, it is reasonable to
assume that the variations of quad-band source positions due to such invisible
structure are on the order of 0.2 -- 1.0\,mas. While actual variations may not
reach that level when comparing positions at different VLBI frequencies (since
the positions at various frequencies are all shifted towards the jet), the
effect of invisible structure will fully show up when comparing to optical
positions, and as such is one of the major causes for optical-radio offsets.
This particular characteristics, although limited to core-dominated
sources, is relevant for an investigation of the impact of source structure
because the majority of the CRF sources are indeed core-dominated.

For cases where the radio emission is jet-dominated, position variations result
from the visible structure instead and can be larger than 1.0\,mas, as we have
seen for 2229+695 in Sect.\,\ref{one_motion} and for 0723$-$008 in
Sect.\,\ref{sec:difference}. This is one of the reasons why the position
variations for sources 3C418 (core-dominated)
and 2229+695 (jet-dominated) are categorized differently. Moreover, in
extreme cases where the peak component interchanges between two components
(e.g., the core and a jet component), a large jump in the
astrometric position can happen in a short period of time, such as that
observed for 3C48, which showed a position change of 60\,mas
\citep{2022MNRAS.512..874T}.

The case of 3C418 highlights the importance of registering the
images across frequencies when modeling source structure in
geodetic observations, whereas the cases of 0723$-$008 and 2229+695 highlight
the importance of registering across time. One cannot simply rely on features
such as the peak
component or the modeled core component at each band to this end.

\section{Conclusions and outlook} \label{sec:cite}

We have analyzed delay observations from 177 quad-band VLBI sessions
coordinated by the IVS from December 2017 to June 2024 and made both
session-wise and global solutions. Based on these data, a catalog comprising
positions for 377 sources was produced. In the session-wise analysis, some
flagged observations were restored after allowing for the estimation of source
positions.
Besides the global solution that led to the quad-band catalog, dedicated
global solution was done in turn for each radio
source where the position of this radio source was estimated session-wise,
which was further used to investigate the source position stability based on
the such-derived position time series.
Three representative sources, 1803+784, 2229+695, and 3C418, were used to study
details of the source position variations and together with insights from radio
images to investigate the impact of visible and invisible structure. By removing
the overall systematic variations in the position time series, the full
capability of quad-band VLBI observations to measure source positions was
assessed. We also compared the quad-band source positions with the S/X
positions in the ICRF3.

From these studies, we draw the following conclusions:
\begin{enumerate}
\item The noise floor of the quad-band source positions has been determined as 
a function of declination. While it
is \,40\,$\mu$as for both RA$^{*}$ and Dec in the Northern sky, it
increases as declination decreases, reaching 200\,$\mu$as for RA$^{*}$ and
500\,$\mu$as for
Dec in the Southern sky. The major factor for such increased uncertainties in
Dec is the limitation of the current antenna network and the resulting limited
observation sensitivity to the declination coordinate.
\item For sources with declination above 20$\degree$, the residuals of the
source coordinates are on the order of 0.1 -- 0.2\,mas (in the
sense of standard deviation), and the dominant cause is
source structure. This limits the
accuracy of quad-band source positions.
In particular, the source positions estimated
from observations over a long time span do not represent the actual positions
at a given epoch.
\item Based on the impact on geodesy and astrometry, source
structure may be divided into two parts: invisible/in-beam structure
(i.e., within the beam size) and visible structure (on larger scales). The
latter mainly causes closure delays leading to large post-fit delay residuals in
geodetic solutions whereas the former causes source position changes.
\item For some sources, such as 0642+449, 1039+811, 1030+415, and 2229+695, the 
brightest component appears to be a moving jet component instead of
the core. This jet component drives the measured source position continuously
in one direction away from the core, with a large magnitude, on the
milli-arcsecond level.
\item The most common position variation is back-and-forth motion
along one direction, with a magnitude that is changing from source to
source. It is caused by the in-beam structure and jet kinematics. Given
that the typical beam widths of quad-band VLBI observations are 0.4 --
2.0\,mas, source position offsets in the range of 0.2 -- 1.0\,mas may result
from such effect. These offsets may not
be fully detected by VLBI observations at the different radio frequencies,
however, they will contribute significantly to the position differences
observed between radio and optical.
\item The median and mean arc lengths between
the quad-band and S/X positions over the 377 sources in our catalog are
0.175\,mas and 0.310\,mas, respectively. About 15\% of the quad-band sources
are found to have significant offsets ($>$ 3$\sigma$), with median and
mean values of 0.58\,mas and 0.82\,mas, respectively.
\item About 6\% of the radio sources have position offsets larger than 0.8\,mas
($>$ 3$\sigma$) with respect to the S/X positions. The
largest offset occurs for the source
0723$-$008, with an arc length of 4.73$\pm$0.29\,mas. Based on its Gaia
position and a MOJAVE image, we suggest that the Gaia and S/X
positions for this source are close to the location of the core, while the
quad-band position is consistent with the position of a jet component
that is brighter than the core.
\item A quad-band source position catalog is mandatory to analyze quad-band
VLBI observations. The use of catalogs at other frequencies for this purpose
would add systematic errors because of potential source position offsets
between the different frequencies.
\end{enumerate}

Because of the impact of in-beam structure on
source positions, moving to higher radio frequencies may have significance
for astrometry and geodesy in the future. Such VLBI
observations have been reported in recent studies
\citep[e.g., ][]{2023AJ....165..139D,2024AJ....168...76P,
2024arXiv240907309X}. This may be overshadowed by other
challenges, however, such as the radio
sources becoming weaker at higher radio frequencies, the atmosphere being less
transparent, and if observing at
a single band, ionospheric delays. To improve the accuracy of future
quad-band CRFs, our plan is (1) to  study, characterize, monitor,
and interpret the frequency- and time-dependent structure of the
AGNs based on VGOS observations and (2) to develop approaches to model them as
point-like fiducial anchors on the sky.

\begin{longrotatetable}
\begin{table}
\begin{center}
\tablenum{3}
\small
\caption{Source position catalog from quad-band VLBI observations.}
\begin{tabular}{c@{\hskip 6pt}l@{}c@{\hskip 2pt}l@{\hskip 2pt}r@{\hskip 
10pt}r@{\hskip 16pt}rr@{}c@{\ \ \ \ \ }r@{\hskip 9pt}ccc@{}c@{\hskip 
10pt}rr@{\hskip 3pt}r}
\hline
\hline
\noalign{\smallskip}
\multicolumn{2}{c}{Source identification}
&&Cat.
&\ \hfill Right ascension\hfill\
&\ \hfill Declination\hfill\
&\multicolumn{2}{l}{Coordinate uncertainty}
&&Correl.
&\multicolumn{3}{c}{Epoch of sessions}
&&\multicolumn{3}{c}{Observations}
\\
\cline{1-2}\cline{7-8}\cline{11-13}\cline{15-17}
\noalign{\smallskip}
ICRF designation &
IERS name &
&&
 (h\ \ m\ \ s)\hfill\ \   &
($\degr$\ \ \ $\arcmin$\ \ \ $\arcsec$)\hfill\ \   &
\ \hfill (s)\hfill\ &
\ \hfill ($\arcsec$)\hfill\ &&
&
Mean &
First &
Last &
&
$N_{\rm ses}$ &
$N_{\rm del}$ &
$N_{\rm rat}$
\\
\noalign{\smallskip}
\hline
\noalign{\smallskip}
(1) & (2) & & (3) & (4) & (5) & (6) &(7) & & (8) & (9) & (10) & (11) & & (12) 
&(13) & (14) \\
\noalign{\smallskip}
\hline
\noalign{\smallskip}
ICRF J000613.8$-$062335 &   0003$-$066  && {\hskip 1pt} &   00 06 13.89288736 &
$-$06 23 35.3356487  &   0.00000244  &   0.0000842&& $-$0.6000&  59628.5 &
58910.0 & 60347.0&&     52&     4994 &  0 \\
ICRF J001708.4+813508 &   0014+813  && {\hskip 1pt} &   00 17 08.47496512 &
+81 35 08.1353152  &   0.00002159  &   0.0000520&&  0.0670&  60305.5 &  60264.0
& 60347.0&&      2 &      366&  0 \\
ICRF J001945.7+732730 &   0016+731  && {\hskip 1pt} &   00 19 45.78630767 &
+73 27 30.0176149  &   0.00000146  &   0.0000060&& $-$0.0130&  59278.0 &
58090.0 & 60466.0&&     82 &   43421&  0 \\
ICRF J001937.8+202145 &   0017+200  && {\hskip 1pt} &   00 19 37.85449129 &
+20 21 45.6444880  &   0.00000085  &   0.0000277&& $-$0.3970&  59649.5 &
58812.0 & 60487.0&&     97&    12768&  0 \\
ICRF J002232.4+060804 &   0019+058  && {\hskip 1pt} &   00 22 32.44120545 &
+06 08 04.2689715  &   0.00000167  &   0.0000637&& $-$0.5220&  59649.5 &
58812.0 & 60487.0&&      87 &    4952&  0 \\
ICRF J002829.8+200026 &   0025+197  &C & {\hskip 1pt} &   00 28 29.81847366 &
+20 00 26.7439590  &   0.00000226  &   0.0000755&& $-$0.2880&  59623.0 &
58857.0 & 60389.0&&     30 &    1658&   0 \\

\hline
\label{Tab:VGOS_coordinates}
\end{tabular}
\tablecomments{Column (1) is the ICRF name. In column (3), character ``C''
stands for the datum sources and ``S'' for the sources with positions estimated
as local parameters. The
positions of the sources in the ``S'' category are the mean of the
session-wise positions. Columns (12) and (13) provide the number of sessions
and the
number of used
observations, respectively. Column (14), which provides the number
of delay rate observations, is kept only for consistency with
the ICRF catalogs. Full table is available in a machine-readable format.}
\end{center}
\end{table}
\end{longrotatetable}

\begin{acknowledgments}
All components of the International VLBI Service for Geodesy and Astrometry 
are deeply appreciated for providing the VLBI observations. This research has 
made use of data from the MOJAVE database that is maintained 
by the MOJAVE team \citep{2018ApJS..234...12L}. The authors would like to
thank Leonid Petrov, Shuangjing Xu, Ezequiel
Albentosa-Ruiz, Claudio Rivera, and Wara Chamani for discussions on this topic.
Comments from an anonymous reviewer enhanced the manuscript
significantly and were greatly appreciated. This research is supported
by the ERC grant no. 101076060.
\end{acknowledgments}

%

\vspace{5mm}
\facilities{VGOS, VLBI, Gaia}


\software{ehtim \citep{2018ApJ...857...23C}, difmap 
\citep{1994BAAS...26..987S}, pSolve
          }



%


\bibliography{vgossrc_R1}{}

\begin{thebibliography}{}
\expandafter\ifx\csname natexlab\endcsname\relax\def\natexlab#1{#1}\fi
\providecommand{\url}[1]{\href{#1}{#1}}
\providecommand{\dodoi}[1]{doi:~\href{http://doi.org/#1}{\nolinkurl{#1}}}
\providecommand{\doeprint}[1]{\href{http://ascl.net/#1}{\nolinkurl{http://ascl.net/#1}}}
\providecommand{\doarXiv}[1]{\href{https://arxiv.org/abs/#1}{\nolinkurl{https://arxiv.org/abs/#1}}}

\bibitem[{{Altamimi} {et~al.}(2023){Altamimi}, {Rebischung}, {Collilieux},
  {M{\'e}tivier}, \& {Chanard}}]{2023JGeod..97...47A}
{Altamimi}, Z., {Rebischung}, P., {Collilieux}, X., {M{\'e}tivier}, L., \&
  {Chanard}, K. 2023, Journal of Geodesy, 97, 47,
  \dodoi{10.1007/s00190-023-01738-w}

\bibitem[{{Altamimi} {et~al.}(2016){Altamimi}, {Rebischung}, {M{\'e}tivier}, \&
  {Collilieux}}]{2016JGRB..121.6109A}
{Altamimi}, Z., {Rebischung}, P., {M{\'e}tivier}, L., \& {Collilieux}, X. 2016,
  Journal of Geophysical Research (Solid Earth), 121, 6109,
  \dodoi{10.1002/2016JB013098}

\bibitem[{{Anderson} \& {Xu}(2018)}]{2018JGRB..12310162A}
{Anderson}, J.~M., \& {Xu}, M.~H. 2018, Journal of Geophysical Research (Solid
  Earth), 123, 10,162, \dodoi{10.1029/2018JB015550}

\bibitem[{{Bizouard} {et~al.}(2019){Bizouard}, {Lambert}, {Gattano}, {Becker},
  \& {Richard}}]{2019JGeod..93..621B}
{Bizouard}, C., {Lambert}, S., {Gattano}, C., {Becker}, O., \& {Richard}, J.-Y.
  2019, Journal of Geodesy, 93, 621, \dodoi{10.1007/s00190-018-1186-3}

\bibitem[{{Chael} {et~al.}(2018){Chael}, {Johnson}, {Bouman}, {Blackburn},
  {Akiyama}, \& {Narayan}}]{2018ApJ...857...23C}
{Chael}, A.~A., {Johnson}, M.~D., {Bouman}, K.~L., {et~al.} 2018, \apj, 857,
  23, \dodoi{10.3847/1538-4357/aab6a8}

\bibitem[{{Charlot} {et~al.}(2020){Charlot}, {Jacobs}, {Gordon}, {Lambert}, {de
  Witt}, {B{\"o}hm}, {Fey}, {Heinkelmann}, {Skurikhina}, {Titov}, {Arias},
  {Bolotin}, {Bourda}, {Ma}, {Malkin}, {Nothnagel}, {Mayer}, {MacMillan},
  {Nilsson}, \& {Gaume}}]{2020A&A...644A.159C}
{Charlot}, P., {Jacobs}, C.~S., {Gordon}, D., {et~al.} 2020, \aap, 644, A159,
  \dodoi{10.1051/0004-6361/202038368}

\bibitem[{{Chen} {et~al.}(2023){Chen}, {Zhang}, {Zhang}, {Yang}, {Xu}, {Sun},
  {Mai}, {Shu}, \& {Wang}}]{2023MNRAS.524.5357C}
{Chen}, W., {Zhang}, B., {Zhang}, J., {et~al.} 2023, \mnras, 524, 5357,
  \dodoi{10.1093/mnras/stad1214}

\bibitem[{{Cigan} {et~al.}(2024){Cigan}, {Makarov}, {Secrest}, {Gordon},
  {Johnson}, \& {Lambert}}]{2024ApJS..274...28C}
{Cigan}, P., {Makarov}, V.~V., {Secrest}, N.~J., {et~al.} 2024, \apjs, 274, 28,
  \dodoi{10.3847/1538-4365/ad6772}

\bibitem[{{de Witt} {et~al.}(2023){de Witt}, {Jacobs}, {Gordon}, {Bietenholz},
  {Nickola}, \& {Bertarini}}]{2023AJ....165..139D}
{de Witt}, A., {Jacobs}, C.~S., {Gordon}, D., {et~al.} 2023, \aj, 165, 139,
  \dodoi{10.3847/1538-3881/aca012}

\bibitem[{{Fey} \& {Charlot}(1997)}]{1997ApJS..111...95F}
{Fey}, A.~L., \& {Charlot}, P. 1997, \apjs, 111, 95, \dodoi{10.1086/313017}

\bibitem[{{Fey} {et~al.}(2015){Fey}, {Gordon}, {Jacobs}, {Ma}, {Gaume},
  {Arias}, {Bianco}, {Boboltz}, {B{\"o}ckmann}, {Bolotin}, {Charlot},
  {Collioud}, {Engelhardt}, {Gipson}, {Gontier}, {Heinkelmann}, {Kurdubov},
  {Lambert}, {Lytvyn}, {MacMillan}, {Malkin}, {Nothnagel}, {Ojha},
  {Skurikhina}, {Sokolova}, {Souchay}, {Sovers}, {Tesmer}, {Titov}, {Wang}, \&
  {Zharov}}]{2015AJ....150...58F}
{Fey}, A.~L., {Gordon}, D., {Jacobs}, C.~S., {et~al.} 2015, \aj, 150, 58,
  \dodoi{10.1088/0004-6256/150/2/58}

\bibitem[{{Gaia Collaboration} {et~al.}(2016){Gaia Collaboration}, {Brown},
  {Vallenari}, {Prusti}, {de Bruijne}, {Mignard}, {Drimmel}, {Babusiaux},
  {Bailer-Jones}, {Bastian}, {Biermann}, {Evans}, {Eyer}, {Jansen}, {Jordi},
  {Katz}, {Klioner}, {Lammers}, {Lindegren}, {Luri}, {O'Mullane}, {Panem},
  {Pourbaix}, {Randich}, {Sartoretti}, {Siddiqui}, {Soubiran}, {Valette}, {van
  Leeuwen}, {Walton}, {Aerts}, {Arenou}, {Cropper}, {H{\o}g}, {Lattanzi},
  {Grebel}, {Holland}, {Huc}, {Passot}, {Perryman}, {Bramante}, {Cacciari},
  {Casta{\~n}eda}, {Chaoul}, {Cheek}, {De Angeli}, {Fabricius}, {Guerra},
  {Hern{\'a}ndez}, {Jean-Antoine-Piccolo}, {Masana}, {Messineo}, {Mowlavi},
  {Nienartowicz}, {Ord{\'o}{\~n}ez-Blanco}, {Panuzzo}, {Portell}, {Richards},
  {Riello}, {Seabroke}, {Tanga}, {Th{\'e}venin}, {Torra}, {Els},
  {Gracia-Abril}, {Comoretto}, {Garcia-Reinaldos}, {Lock}, {Mercier},
  {Altmann}, {Andrae}, {Astraatmadja}, {Bellas-Velidis}, {Benson}, {Berthier},
  {Blomme}, {Busso}, {Carry}, {Cellino}, {Clementini}, {Cowell}, {Creevey},
  {Cuypers}, {Davidson}, {De Ridder}, {de Torres}, {Delchambre}, {Dell'Oro},
  {Ducourant}, {Fr{\'e}mat}, {Garc{\'\i}a-Torres}, {Gosset}, {Halbwachs},
  {Hambly}, {Harrison}, {Hauser}, {Hestroffer}, {Hodgkin}, {Huckle}, {Hutton},
  {Jasniewicz}, {Jordan}, {Kontizas}, {Korn}, {Lanzafame}, {Manteiga},
  {Moitinho}, {Muinonen}, {Osinde}, {Pancino}, {Pauwels}, {Petit},
  {Recio-Blanco}, {Robin}, {Sarro}, {Siopis}, {Smith}, {Smith}, {Sozzetti},
  {Thuillot}, {van Reeven}, {Viala}, {Abbas}, {Abreu Aramburu}, {Accart},
  {Aguado}, {Allan}, {Allasia}, {Altavilla}, {{\'A}lvarez}, {Alves},
  {Anderson}, {Andrei}, {Anglada Varela}, {Antiche}, {Antoja}, {Ant{\'o}n},
  {Arcay}, {Bach}, {Baker}, {Balaguer-N{\'u}{\~n}ez}, {Barache}, {Barata},
  {Barbier}, {Barblan}, {Barrado y Navascu{\'e}s}, {Barros}, {Barstow},
  {Becciani}, {Bellazzini}, {Bello Garc{\'\i}a}, {Belokurov}, {Bendjoya},
  {Berihuete}, {Bianchi}, {Bienaym{\'e}}, {Billebaud}, {Blagorodnova},
  {Blanco-Cuaresma}, {Boch}, {Bombrun}, {Borrachero}, {Bouquillon}, {Bourda},
  {Bouy}, {Bragaglia}, {Breddels}, {Brouillet}, {Br{\"u}semeister},
  {Bucciarelli}, {Burgess}, {Burgon}, {Burlacu}, {Busonero}, {Buzzi}, {Caffau},
  {Cambras}, {Campbell}, {Cancelliere}, {Cantat-Gaudin}, {Carlucci},
  {Carrasco}, {Castellani}, {Charlot}, {Charnas}, {Chiavassa}, {Clotet},
  {Cocozza}, {Collins}, {Costigan}, {Crifo}, {Cross}, {Crosta}, {Crowley},
  {Dafonte}, {Damerdji}, {Dapergolas}, {David}, {David}, {De Cat}, {de Felice},
  {de Laverny}, {De Luise}, {De March}, {de Martino}, {de Souza}, {Debosscher},
  {del Pozo}, {Delbo}, {Delgado}, {Delgado}, {Di Matteo}, {Diakite},
  {Distefano}, {Dolding}, {Dos Anjos}, {Drazinos}, {Duran}, {Dzigan},
  {Edvardsson}, {Enke}, {Evans}, {Eynard Bontemps}, {Fabre}, {Fabrizio},
  {Faigler}, {Falc{\~a}o}, {Farr{\`a}s Casas}, {Federici}, {Fedorets},
  {Fern{\'a}ndez-Hern{\'a}ndez}, {Fernique}, {Fienga}, {Figueras}, {Filippi},
  {Findeisen}, {Fonti}, {Fouesneau}, {Fraile}, {Fraser}, {Fuchs}, {Gai},
  {Galleti}, {Galluccio}, {Garabato}, {Garc{\'\i}a-Sedano}, {Garofalo},
  {Garralda}, {Gavras}, {Gerssen}, {Geyer}, {Gilmore}, {Girona}, {Giuffrida},
  {Gomes}, {Gonz{\'a}lez-Marcos}, {Gonz{\'a}lez-N{\'u}{\~n}ez},
  {Gonz{\'a}lez-Vidal}, {Granvik}, {Guerrier}, {Guillout}, {Guiraud},
  {G{\'u}rpide}, {Guti{\'e}rrez-S{\'a}nchez}, {Guy}, {Haigron},
  {Hatzidimitriou}, {Haywood}, {Heiter}, {Helmi}, {Hobbs}, {Hofmann}, {Holl},
  {Holland}, {Hunt}, {Hypki}, {Icardi}, {Irwin}, {Jevardat de Fombelle},
  {Jofr{\'e}}, {Jonker}, {Jorissen}, {Julbe}, {Karampelas}, {Kochoska},
  {Kohley}, {Kolenberg}, {Kontizas}, {Koposov}, {Kordopatis}, {Koubsky},
  {Krone-Martins}, {Kudryashova}, {Kull}, {Bachchan}, {Lacoste-Seris}, {Lanza},
  {Lavigne}, {Le Poncin-Lafitte}, {Lebreton}, {Lebzelter}, {Leccia}, {Leclerc},
  {Lecoeur-Taibi}, {Lemaitre}, {Lenhardt}, {Leroux}, {Liao}, {Licata},
  {Lindstr{\o}m}, {Lister}, {Livanou}, {Lobel}, {L{\"o}ffler}, {L{\'o}pez},
  {Lorenz}, {MacDonald}, {Magalh{\~a}es Fernandes}, {Managau}, {Mann},
  {Mantelet}, {Marchal}, {Marchant}, {Marconi}, {Marinoni}, {Marrese},
  {Marschalk{\'o}}, {Marshall}, {Mart{\'\i}n-Fleitas}, {Martino}, {Mary},
  {Matijevi{\v{c}}}, {Mazeh}, {McMillan}, {Messina}, {Michalik}, {Millar},
  {Miranda}, {Molina}, {Molinaro}, {Molinaro}, {Moln{\'a}r}, {Moniez},
  {Montegriffo}, {Mor}, {Mora}, {Morbidelli}, {Morel}, {Morgenthaler},
  {Morris}, {Mulone}, {Muraveva}, {Musella}, {Narbonne}, {Nelemans},
  {Nicastro}, {Noval}, {Ord{\'e}novic}, {Ordieres-Mer{\'e}}, {Osborne},
  {Pagani}, {Pagano}, {Pailler}, {Palacin}, {Palaversa}, {Parsons}, {Pecoraro},
  {Pedrosa}, {Pentik{\"a}inen}, {Pichon}, {Piersimoni}, {Pineau}, {Plachy},
  {Plum}, {Poujoulet}, {Pr{\v{s}}a}, {Pulone}, {Ragaini}, {Rago}, {Rambaux},
  {Ramos-Lerate}, {Ranalli}, {Rauw}, {Read}, {Regibo}, {Reyl{\'e}}, {Ribeiro},
  {Rimoldini}, {Ripepi}, {Riva}, {Rixon}, {Roelens}, {Romero-G{\'o}mez},
  {Rowell}, {Royer}, {Ruiz-Dern}, {Sadowski}, {Sagrist{\`a} Sell{\'e}s},
  {Sahlmann}, {Salgado}, {Salguero}, {Sarasso}, {Savietto}, {Schultheis},
  {Sciacca}, {Segol}, {Segovia}, {Segransan}, {Shih}, {Smareglia}, {Smart},
  {Solano}, {Solitro}, {Sordo}, {Soria Nieto}, {Souchay}, {Spagna}, {Spoto},
  {Stampa}, {Steele}, {Steidelm{\"u}ller}, {Stephenson}, {Stoev}, {Suess},
  {S{\"u}veges}, {Surdej}, {Szabados}, {Szegedi-Elek}, {Tapiador}, {Taris},
  {Tauran}, {Taylor}, {Teixeira}, {Terrett}, {Tingley}, {Trager}, {Turon},
  {Ulla}, {Utrilla}, {Valentini}, {van Elteren}, {Van Hemelryck}, {van
  Leeuwen}, {Varadi}, {Vecchiato}, {Veljanoski}, {Via}, {Vicente}, {Vogt},
  {Voss}, {Votruba}, {Voutsinas}, {Walmsley}, {Weiler}, {Weingrill}, {Wevers},
  {Wyrzykowski}, {Yoldas}, {{\v{Z}}erjal}, {Zucker}, {Zurbach}, {Zwitter},
  {Alecu}, {Allen}, {Allende Prieto}, {Amorim}, {Anglada-Escud{\'e}},
  {Arsenijevic}, {Azaz}, {Balm}, {Beck}, {Bernstein}, {Bigot}, {Bijaoui},
  {Blasco}, {Bonfigli}, {Bono}, {Boudreault}, {Bressan}, {Brown}, {Brunet},
  {Bunclark}, {Buonanno}, {Butkevich}, {Carret}, {Carrion}, {Chemin},
  {Ch{\'e}reau}, {Corcione}, {Darmigny}, {de Boer}, {de Teodoro}, {de Zeeuw},
  {Delle Luche}, {Domingues}, {Dubath}, {Fodor}, {Fr{\'e}zouls}, {Fries},
  {Fustes}, {Fyfe}, {Gallardo}, {Gallegos}, {Gardiol}, {Gebran}, {Gomboc},
  {G{\'o}mez}, {Grux}, {Gueguen}, {Heyrovsky}, {Hoar}, {Iannicola}, {Isasi
  Parache}, {Janotto}, {Joliet}, {Jonckheere}, {Keil}, {Kim}, {Klagyivik},
  {Klar}, {Knude}, {Kochukhov}, {Kolka}, {Kos}, {Kutka}, {Lainey}, {LeBouquin},
  {Liu}, {Loreggia}, {Makarov}, {Marseille}, {Martayan}, {Martinez-Rubi},
  {Massart}, {Meynadier}, {Mignot}, {Munari}, {Nguyen}, {Nordlander}, {Ocvirk},
  {O'Flaherty}, {Olias Sanz}, {Ortiz}, {Osorio}, {Oszkiewicz}, {Ouzounis},
  {Palmer}, {Park}, {Pasquato}, {Peltzer}, {Peralta}, {P{\'e}turaud},
  {Pieniluoma}, {Pigozzi}, {Poels}, {Prat}, {Prod'homme}, {Raison}, {Rebordao},
  {Risquez}, {Rocca-Volmerange}, {Rosen}, {Ruiz-Fuertes}, {Russo}, {Sembay},
  {Serraller Vizcaino}, {Short}, {Siebert}, {Silva}, {Sinachopoulos}, {Slezak},
  {Soffel}, {Sosnowska}, {Strai{\v{z}}ys}, {ter Linden}, {Terrell}, {Theil},
  {Tiede}, {Troisi}, {Tsalmantza}, {Tur}, {Vaccari}, {Vachier}, {Valles}, {Van
  Hamme}, {Veltz}, {Virtanen}, {Wallut}, {Wichmann}, {Wilkinson}, {Ziaeepour},
  \& {Zschocke}}]{2016A&A...595A...2G}
{Gaia Collaboration}, {Brown}, A.~G.~A., {Vallenari}, A., {et~al.} 2016, \aap,
  595, A2, \dodoi{10.1051/0004-6361/201629512}

\bibitem[{{Gaia Collaboration} {et~al.}(2021){Gaia Collaboration}, {Klioner},
  {Mignard}, {Lindegren}, {Bastian}, {McMillan}, {Hern{\'a}ndez}, {Hobbs},
  {Ramos-Lerate}, {Biermann}, {Bombrun}, {de Torres}, {Gerlach}, {Geyer},
  {Hilger}, {Lammers}, {Steidelm{\"u}ller}, {Stephenson}, {Brown}, {Vallenari},
  {Prusti}, {de Bruijne}, {Babusiaux}, {Creevey}, {Evans}, {Eyer}, {Hutton},
  {Jansen}, {Jordi}, {Luri}, {Panem}, {Pourbaix}, {Randich}, {Sartoretti},
  {Soubiran}, {Walton}, {Arenou}, {Bailer-Jones}, {Cropper}, {Drimmel}, {Katz},
  {Lattanzi}, {van Leeuwen}, {Bakker}, {Casta{\~n}eda}, {De Angeli},
  {Ducourant}, {Fabricius}, {Fouesneau}, {Fr{\'e}mat}, {Guerra}, {Guerrier},
  {Guiraud}, {Jean-Antoine Piccolo}, {Masana}, {Messineo}, {Mowlavi},
  {Nicolas}, {Nienartowicz}, {Pailler}, {Panuzzo}, {Riclet}, {Roux},
  {Seabroke}, {Sordo}, {Tanga}, {Th{\'e}venin}, {Gracia-Abril}, {Portell},
  {Teyssier}, {Altmann}, {Andrae}, {Bellas-Velidis}, {Benson}, {Berthier},
  {Blomme}, {Brugaletta}, {Burgess}, {Busso}, {Carry}, {Cellino}, {Cheek},
  {Clementini}, {Damerdji}, {Davidson}, {Delchambre}, {Dell'Oro},
  {Fern{\'a}ndez-Hern{\'a}ndez}, {Galluccio}, {Garc{\'\i}a-Lario},
  {Garcia-Reinaldos}, {Gonz{\'a}lez-N{\'u}{\~n}ez}, {Gosset}, {Haigron},
  {Halbwachs}, {Hambly}, {Harrison}, {Hatzidimitriou}, {Heiter}, {Hestroffer},
  {Hodgkin}, {Holl}, {Jan{\ss}en}, {Jevardat de Fombelle}, {Jordan},
  {Krone-Martins}, {Lanzafame}, {L{\"o}ffler}, {Lorca}, {Manteiga}, {Marchal},
  {Marrese}, {Moitinho}, {Mora}, {Muinonen}, {Osborne}, {Pancino}, {Pauwels},
  {Recio-Blanco}, {Richards}, {Riello}, {Rimoldini}, {Robin}, {Roegiers},
  {Rybizki}, {Sarro}, {Siopis}, {Smith}, {Sozzetti}, {Ulla}, {Utrilla}, {van
  Leeuwen}, {van Reeven}, {Abbas}, {Abreu Aramburu}, {Accart}, {Aerts},
  {Aguado}, {Ajaj}, {Altavilla}, {{\'A}lvarez}, {{\'A}lvarez Cid-Fuentes},
  {Alves}, {Anderson}, {Anglada Varela}, {Antoja}, {Audard}, {Baines}, {Baker},
  {Balaguer-N{\'u}{\~n}ez}, {Balbinot}, {Balog}, {Barache}, {Barbato},
  {Barros}, {Barstow}, {Bartolom{\'e}}, {Bassilana}, {Bauchet},
  {Baudesson-Stella}, {Becciani}, {Bellazzini}, {Bernet}, {Bertone}, {Bianchi},
  {Blanco-Cuaresma}, {Boch}, {Bossini}, {Bouquillon}, {Bramante}, {Breedt},
  {Bressan}, {Brouillet}, {Bucciarelli}, {Burlacu}, {Busonero}, {Butkevich},
  {Buzzi}, {Caffau}, {Cancelliere}, {C{\'a}novas}, {Cantat-Gaudin}, {Carballo},
  {Carlucci}, {Carnerero}, {Carrasco}, {Casamiquela}, {Castellani},
  {Castro-Ginard}, {Castro Sampol}, {Chaoul}, {Charlot}, {Chemin}, {Chiavassa},
  {Comoretto}, {Cooper}, {Cornez}, {Cowell}, {Crifo}, {Crosta}, {Crowley},
  {Dafonte}, {Dapergolas}, {David}, {David}, {de Laverny}, {De Luise}, {De
  March}, {De Ridder}, {de Souza}, {de Teodoro}, {del Peloso}, {del Pozo},
  {Delgado}, {Delgado}, {Delisle}, {Di Matteo}, {Diakite}, {Diener},
  {Distefano}, {Dolding}, {Eappachen}, {Enke}, {Esquej}, {Fabre}, {Fabrizio},
  {Faigler}, {Fedorets}, {Fernique}, {Fienga}, {Figueras}, {Fouron},
  {Fragkoudi}, {Fraile}, {Franke}, {Gai}, {Garabato}, {Garcia-Gutierrez},
  {Garc{\'\i}a-Torres}, {Garofalo}, {Gavras}, {Giacobbe}, {Gilmore}, {Girona},
  {Giuffrida}, {Gomez}, {Gonzalez-Santamaria}, {Gonz{\'a}lez-Vidal}, {Granvik},
  {Guti{\'e}rrez-S{\'a}nchez}, {Guy}, {Hauser}, {Haywood}, {Helmi}, {Hidalgo},
  {H{\l}adczuk}, {Holland}, {Huckle}, {Jasniewicz}, {Jonker}, {Juaristi
  Campillo}, {Julbe}, {Karbevska}, {Kervella}, {Khanna}, {Kochoska},
  {Kordopatis}, {Korn}, {Kostrzewa-Rutkowska}, {Kruszy{\'n}ska}, {Lambert},
  {Lanza}, {Lasne}, {Le Campion}, {Le Fustec}, {Lebreton}, {Lebzelter},
  {Leccia}, {Leclerc}, {Lecoeur-Taibi}, {Liao}, {Licata}, {Lindstr{\o}m},
  {Lister}, {Livanou}, {Lobel}, {Madrero Pardo}, {Managau}, {Mann}, {Marchant},
  {Marconi}, {Marcos Santos}, {Marinoni}, {Marocco}, {Marshall}, {Martin Polo},
  {Mart{\'\i}n-Fleitas}, {Masip}, {Massari}, {Mastrobuono-Battisti}, {Mazeh},
  {Messina}, {Michalik}, {Millar}, {Mints}, {Molina}, {Molinaro}, {Moln{\'a}r},
  {Montegriffo}, {Mor}, {Morbidelli}, {Morel}, {Morris}, {Mulone}, {Munoz},
  {Muraveva}, {Murphy}, {Musella}, {Noval}, {Ord{\'e}novic}, {Orr{\`u}},
  {Osinde}, {Pagani}, {Pagano}, {Palaversa}, {Palicio}, {Panahi}, {Pawlak},
  {Pe{\~n}alosa Esteller}, {Penttil{\"a}}, {Piersimoni}, {Pineau}, {Plachy},
  {Plum}, {Poggio}, {Poretti}, {Poujoulet}, {Pr{\v{s}}a}, {Pulone}, {Racero},
  {Ragaini}, {Rainer}, {Raiteri}, {Rambaux}, {Ramos}, {Re Fiorentin}, {Regibo},
  {Reyl{\'e}}, {Ripepi}, {Riva}, {Rixon}, {Robichon}, {Robin}, {Roelens},
  {Rohrbasser}, {Romero-G{\'o}mez}, {Rowell}, {Royer}, {Rybicki}, {Sadowski},
  {Sagrist{\`a} Sell{\'e}s}, {Sahlmann}, {Salgado}, {Salguero}, {Samaras},
  {Sanchez Gimenez}, {Sanna}, {Santove{\~n}a}, {Sarasso}, {Schultheis},
  {Sciacca}, {Segol}, {Segovia}, {S{\'e}gransan}, {Semeux}, {Siddiqui},
  {Siebert}, {Siltala}, {Slezak}, {Smart}, {Solano}, {Solitro}, {Souami},
  {Souchay}, {Spagna}, {Spoto}, {Steele}, {S{\"u}veges}, {Szabados},
  {Szegedi-Elek}, {Taris}, {Tauran}, {Taylor}, {Teixeira}, {Thuillot},
  {Tonello}, {Torra}, {Torra}, {Turon}, {Unger}, {Vaillant}, {van Dillen},
  {Vanel}, {Vecchiato}, {Viala}, {Vicente}, {Voutsinas}, {Weiler}, {Wevers},
  {Wyrzykowski}, {Yoldas}, {Yvard}, {Zhao}, {Zorec}, {Zucker}, {Zurbach}, \&
  {Zwitter}}]{2021A&A...649A...9G}
{Gaia Collaboration}, {Klioner}, S.~A., {Mignard}, F., {et~al.} 2021, \aap,
  649, A9, \dodoi{10.1051/0004-6361/202039734}

\bibitem[{{Gaia Collaboration} {et~al.}(2022){Gaia Collaboration}, {Klioner},
  {Lindegren}, {Mignard}, {Hern{\'a}ndez}, {Ramos-Lerate}, {Bastian},
  {Biermann}, {Bombrun}, {de Torres}, {Gerlach}, {Geyer}, {Hilger}, {Hobbs},
  {Lammers}, {McMillan}, {Steidelm{\"u}ller}, {Teyssier}, {Raiteri},
  {Bartolom{\'e}}, {Bernet}, {Casta{\~n}eda}, {Clotet}, {Davidson},
  {Fabricius}, {Garralda Torres}, {Gonz{\'a}lez-Vidal}, {Portell}, {Rowell},
  {Torra}, {Torra}, {Brown}, {Vallenari}, {Prusti}, {de Bruijne}, {Arenou},
  {Babusiaux}, {Creevey}, {Ducourant}, {Evans}, {Eyer}, {Guerra}, {Hutton},
  {Jordi}, {Luri}, {Panem}, {Pourbaix}, {Randich}, {Sartoretti}, {Soubiran},
  {Tanga}, {Walton}, {Bailer-Jones}, {Drimmel}, {Jansen}, {Katz}, {Lattanzi},
  {van Leeuwen}, {Bakker}, {Cacciari}, {De Angeli}, {Fouesneau}, {Fr{\'e}mat},
  {Galluccio}, {Guerrier}, {Heiter}, {Masana}, {Messineo}, {Mowlavi},
  {Nicolas}, {Nienartowicz}, {Pailler}, {Panuzzo}, {Riclet}, {Roux},
  {Seabroke}, {Sordo}, {Th{\'e}venin}, {Gracia-Abril}, {Altmann}, {Andrae},
  {Audard}, {Bellas-Velidis}, {Benson}, {Berthier}, {Blomme}, {Burgess},
  {Busonero}, {Busso}, {C{\'a}novas}, {Carry}, {Cellino}, {Cheek},
  {Clementini}, {Damerdji}, {de Teodoro}, {Nu{\~n}ez Campos}, {Delchambre},
  {Dell'Oro}, {Esquej}, {Fern{\'a}ndez-Hern{\'a}ndez}, {Fraile}, {Garabato},
  {Garc{\'\i}a-Lario}, {Gosset}, {Haigron}, {Halbwachs}, {Hambly}, {Harrison},
  {Hestroffer}, {Hodgkin}, {Holl}, {Jan{\ss}en}, {Jevardat de Fombelle},
  {Jordan}, {Krone-Martins}, {Lanzafame}, {L{\"o}ffler}, {Marchal}, {Marrese},
  {Moitinho}, {Muinonen}, {Osborne}, {Pancino}, {Pauwels}, {Recio-Blanco},
  {Reyl{\'e}}, {Riello}, {Rimoldini}, {Roegiers}, {Rybizki}, {Sarro}, {Siopis},
  {Smith}, {Sozzetti}, {Utrilla}, {van Leeuwen}, {Abbas}, {{\'A}brah{\'a}m},
  {Abreu Aramburu}, {Aerts}, {Aguado}, {Ajaj}, {Aldea-Montero}, {Altavilla},
  {{\'A}lvarez}, {Alves}, {Anderson}, {Anglada Varela}, {Antoja}, {Baines},
  {Baker}, {Balaguer-N{\'u}{\~n}ez}, {Balbinot}, {Balog}, {Barache}, {Barbato},
  {Barros}, {Barstow}, {Bassilana}, {Bauchet}, {Becciani}, {Bellazzini},
  {Berihuete}, {Bertone}, {Bianchi}, {Binnenfeld}, {Blanco-Cuaresma}, {Boch},
  {Bossini}, {Bouquillon}, {Bragaglia}, {Bramante}, {Breedt}, {Bressan},
  {Brouillet}, {Brugaletta}, {Bucciarelli}, {Burlacu}, {Butkevich}, {Buzzi},
  {Caffau}, {Cancelliere}, {Cantat-Gaudin}, {Carballo}, {Carlucci},
  {Carnerero}, {Carrasco}, {Casamiquela}, {Castellani}, {Castro-Ginard},
  {Chaoul}, {Charlot}, {Chemin}, {Chiaramida}, {Chiavassa}, {Chornay},
  {Comoretto}, {Contursi}, {Cooper}, {Cornez}, {Cowell}, {Crifo}, {Cropper},
  {Crosta}, {Crowley}, {Dafonte}, {Dapergolas}, {David}, {de Laverny}, {De
  Luise}, {De March}, {De Ridder}, {de Souza}, {del Peloso}, {del Pozo},
  {Delbo}, {Delgado}, {Delisle}, {Demouchy}, {Dharmawardena}, {Diakite},
  {Diener}, {Distefano}, {Dolding}, {Enke}, {Fabre}, {Fabrizio}, {Faigler},
  {Fedorets}, {Fernique}, {Fienga}, {Figueras}, {Fournier}, {Fouron},
  {Fragkoudi}, {Gai}, {Garcia-Gutierrez}, {Garcia-Reinaldos},
  {Garc{\'\i}a-Torres}, {Garofalo}, {Gavel}, {Gavras}, {Giacobbe}, {Gilmore},
  {Girona}, {Giuffrida}, {Gomel}, {Gomez}, {Gonz{\'a}lez-N{\'u}{\~n}ez},
  {Gonz{\'a}lez-Santamar{\'\i}a}, {Granvik}, {Guillout}, {Guiraud},
  {Guti{\'e}rrez-S{\'a}nchez}, {Guy}, {Hatzidimitriou}, {Hauser}, {Haywood},
  {Helmer}, {Helmi}, {Sarmiento}, {Hidalgo}, {H{\l}adczuk}, {Holland},
  {Huckle}, {Jardine}, {Jasniewicz}, {Jean-Antoine Piccolo},
  {Jim{\'e}nez-Arranz}, {Juaristi Campillo}, {Julbe}, {Karbevska}, {Kervella},
  {Khanna}, {Kordopatis}, {Korn}, {K{\'o}sp{\'a}l}, {Kostrzewa-Rutkowska},
  {Kruszy{\'n}ska}, {Kun}, {Laizeau}, {Lambert}, {Lanza}, {Lasne}, {Le
  Campion}, {Lebreton}, {Lebzelter}, {Leccia}, {Leclerc}, {Lecoeur-Taibi},
  {Liao}, {Licata}, {Lindstr{\o}m}, {Lister}, {Livanou}, {Lobel}, {Lorca},
  {Loup}, {Madrero Pardo}, {Magdaleno Romeo}, {Managau}, {Mann}, {Manteiga},
  {Marchant}, {Marconi}, {Marcos}, {Santos}, {Mar{\'\i}n Pina}, {Marinoni},
  {Marocco}, {Marshall}, {Polo}, {Mart{\'\i}n-Fleitas}, {Marton}, {Mary},
  {Masip}, {Massari}, {Mastrobuono-Battisti}, {Mazeh}, {Messina}, {Michalik},
  {Millar}, {Mints}, {Molina}, {Molinaro}, {Moln{\'a}r}, {Monari},
  {Mongui{\'o}}, {Montegriffo}, {Montero}, {Mor}, {Mora}, {Morbidelli},
  {Morel}, {Morris}, {Muraveva}, {Murphy}, {Musella}, {Nagy}, {Noval},
  {Oca{\~n}a}, {Ogden}, {Ordenovic}, {Osinde}, {Pagani}, {Pagano}, {Palaversa},
  {Palicio}, {Pallas-Quintela}, {Panahi}, {Payne-Wardenaar}, {Pe{\~n}alosa
  Esteller}, {Penttil{\"a}}, {Pichon}, {Piersimoni}, {Pineau}, {Plachy},
  {Plum}, {Poggio}, {Pr{\v{s}}a}, {Pulone}, {Racero}, {Ragaini}, {Rainer},
  {Rambaux}, {Ramos}, {Re Fiorentin}, {Regibo}, {Richards}, {Diaz}, {Ripepi},
  {Riva}, {Rix}, {Rixon}, {Robichon}, {Robin}, {Robin}, {Roelens}, {Rogues},
  {Rohrbasser}, {Romero-G{\'o}mez}, {Royer}, {Ruz Mieres}, {Rybicki},
  {Sadowski}, {S{\'a}ez N{\'u}{\~n}ez}, {Sagrist{\`a} Sell{\'e}s}, {Sahlmann},
  {Salguero}, {Samaras}, {Sanchez Gimenez}, {Sanna}, {Santove{\~n}a},
  {Sarasso}, {Schultheis}, {Sciacca}, {Segol}, {Segovia}, {S{\'e}gransan},
  {Semeux}, {Shahaf}, {Siddiqui}, {Siebert}, {Siltala}, {Silvelo}, {Slezak},
  {Slezak}, {Smart}, {Snaith}, {Solano}, {Solitro}, {Souami}, {Souchay},
  {Spagna}, {Spina}, {Spoto}, {Steele}, {Stephenson}, {S{\"u}veges}, {Surdej},
  {Szabados}, {Szegedi-Elek}, {Taris}, {Taylor}, {Teixeira}, {Tolomei},
  {Tonello}, {Torralba Elipe}, {Trabucchi}, {Tsounis}, {Turon}, {Ulla},
  {Unger}, {Vaillant}, {van Dillen}, {van Reeven}, {Vanel}, {Vecchiato},
  {Viala}, {Vicente}, {Voutsinas}, {Weiler}, {Wevers}, {Wyrzykowski}, {Yoldas},
  {Yvard}, {Zhao}, {Zorec}, {Zucker}, \& {Zwitter}}]{2022A&A...667A.148G}
{Gaia Collaboration}, {Klioner}, S.~A., {Lindegren}, L., {et~al.} 2022, \aap,
  667, A148, \dodoi{10.1051/0004-6361/202243483}

\bibitem[{{Gattano} \& {Charlot}(2021)}]{2021A&A...648A.125G}
{Gattano}, C., \& {Charlot}, P. 2021, \aap, 648, A125,
  \dodoi{10.1051/0004-6361/202140377}

\bibitem[{{Karbon} {et~al.}(2017){Karbon}, {Heinkelmann}, {Mora-Diaz}, {Xu},
  {Nilsson}, \& {Schuh}}]{2017JGeod..91..755K}
{Karbon}, M., {Heinkelmann}, R., {Mora-Diaz}, J., {et~al.} 2017, Journal of
  Geodesy, 91, 755, \dodoi{10.1007/s00190-016-0954-1}

\bibitem[{{Kareinen} {et~al.}(2024){Kareinen}, {Zubko}, {Savolainen}, {Xu}, \&
  {Poutanen}}]{2024JGeod..98...38K}
{Kareinen}, N., {Zubko}, N., {Savolainen}, T., {Xu}, M.~H., \& {Poutanen}, M.
  2024, Journal of Geodesy, 98, 38, \dodoi{10.1007/s00190-024-01837-2}

\bibitem[{{Kern} {et~al.}(2023){Kern}, {Schartner}, {B{\"o}hm}, {B{\"o}hm},
  {Nothnagel}, \& {Soja}}]{2023JGeod..97...97K}
{Kern}, L., {Schartner}, M., {B{\"o}hm}, J., {et~al.} 2023, Journal of Geodesy,
  97, 97, \dodoi{10.1007/s00190-023-01792-4}

\bibitem[{{Kovalev} {et~al.}(2008){Kovalev}, {Lobanov}, {Pushkarev}, \&
  {Zensus}}]{2008A&A...483..759K}
{Kovalev}, Y.~Y., {Lobanov}, A.~P., {Pushkarev}, A.~B., \& {Zensus}, J.~A.
  2008, \aap, 483, 759, \dodoi{10.1051/0004-6361:20078679}

\bibitem[{{Kovalev} {et~al.}(2017){Kovalev}, {Petrov}, \&
  {Plavin}}]{2017A&A...598L...1K}
{Kovalev}, Y.~Y., {Petrov}, L., \& {Plavin}, A.~V. 2017, \aap, 598, L1,
  \dodoi{10.1051/0004-6361/201630031}

\bibitem[{{Lambert} \& {Secrest}(2024)}]{2024A&A...684A..93L}
{Lambert}, S., \& {Secrest}, N.~J. 2024, \aap, 684, A93,
  \dodoi{10.1051/0004-6361/202348842}

\bibitem[{{Lambert} {et~al.}(2024){Lambert}, {Sol}, \&
  {Pierron}}]{2024A&A...684A.202L}
{Lambert}, S., {Sol}, H., \& {Pierron}, A. 2024, \aap, 684, A202,
  \dodoi{10.1051/0004-6361/202347210}

\bibitem[{{Lindegren}(2020)}]{2020A&A...633A...1L}
{Lindegren}, L. 2020, \aap, 633, A1, \dodoi{10.1051/0004-6361/201936161}

\bibitem[{{Lister} {et~al.}(2018){Lister}, {Aller}, {Aller}, {Hodge}, {Homan},
  {Kovalev}, {Pushkarev}, \& {Savolainen}}]{2018ApJS..234...12L}
{Lister}, M.~L., {Aller}, M.~F., {Aller}, H.~D., {et~al.} 2018, \apjs, 234, 12,
  \dodoi{10.3847/1538-4365/aa9c44}

\bibitem[{{Liu} {et~al.}(2021){Liu}, {Lambert}, {Charlot}, {Zhu}, {Liu},
  {Jiang}, {Wan}, \& {Ding}}]{2021A&A...652A..87L}
{Liu}, N., {Lambert}, S.~B., {Charlot}, P., {et~al.} 2021, \aap, 652, A87,
  \dodoi{10.1051/0004-6361/202038179}

\bibitem[{{Liu} {et~al.}(2020){Liu}, {Lambert}, {Zhu}, \&
  {Liu}}]{2020A&A...634A..28L}
{Liu}, N., {Lambert}, S.~B., {Zhu}, Z., \& {Liu}, J.~C. 2020, \aap, 634, A28,
  \dodoi{10.1051/0004-6361/201936996}

\bibitem[{{Lunz} {et~al.}(2023){Lunz}, {Anderson}, {Xu}, {Titov},
  {Heinkelmann}, {Johnson}, \& {Schuh}}]{2023A&A...676A..11L}
{Lunz}, S., {Anderson}, J.~M., {Xu}, M.~H., {et~al.} 2023, \aap, 676, A11,
  \dodoi{10.1051/0004-6361/202040266}

\bibitem[{{Lunz} {et~al.}(2024){Lunz}, {Anderson}, {Xu}, {Heinkelmann},
  {Titov}, {Lestrade}, {Johnson}, {Shu}, {Chen}, {Melnikov}, {Mikhailov},
  {McCallum}, {Lopez}, {de Vicente Abad}, \& {Schuh}}]{2024A&A...689..A134}
---. 2024, \aap, 689, A134, \dodoi{10.1051/0004-6361/202142081}

\bibitem[{{Ma} {et~al.}(1998){Ma}, {Arias}, {Eubanks}, {Fey}, {Gontier},
  {Jacobs}, {Sovers}, {Archinal}, \& {Charlot}}]{1998AJ....116..516M}
{Ma}, C., {Arias}, E.~F., {Eubanks}, T.~M., {et~al.} 1998, \aj, 116, 516,
  \dodoi{10.1086/300408}

\bibitem[{{Niell} {et~al.}(2018){Niell}, {Barrett}, {Burns}, {Cappallo},
  {Corey}, {Derome}, {Eckert}, {Elosegui}, {McWhirter}, {Poirier},
  {Rajagopalan}, {Rogers}, {Ruszczyk}, {SooHoo}, {Titus}, {Whitney}, {Behrend},
  {Bolotin}, {Gipson}, {Gordon}, {Himwich}, \&
  {Petrachenko}}]{2018RaSc...53.1269N}
{Niell}, A., {Barrett}, J., {Burns}, A., {et~al.} 2018, Radio Science, 53,
  1269, \dodoi{10.1029/2018RS006617}

\bibitem[{{Petit} \& {Luzum}(2010)}]{2010ITN....36....1P}
{Petit}, G., \& {Luzum}, B. 2010, IERS Technical Note, 36, 1

\bibitem[{{Petrachenko} {et~al.}(2009){Petrachenko}, {Niell}, {Behrend},
  {Corey}, {Boehm}, {Charlot}, {Collioud}, {Gipson}, {Haas}, {Hobiger},
  {Koyama}, {MacMillan}, {Malkin}, {Nilsson}, {Pany}, {Tuccari}, {Whitney}, \&
  {Wresnik}}]{2009vlbi.rept....1P}
{Petrachenko}, B., {Niell}, A., {Behrend}, D., {et~al.} 2009, {Design Aspects
  of the VLBI2010 System. Progress Report of the IVS VLBI2010 Committee, June
  2009.}, NASA/TM-2009-214180, 2009, 62 pages

\bibitem[{{Petrov}(2023)}]{2023AJ....165..183P}
{Petrov}, L. 2023, \aj, 165, 183, \dodoi{10.3847/1538-3881/acc174}

\bibitem[{{Petrov}(2024)}]{2024AJ....168...76P}
---. 2024, \aj, 168, 76, \dodoi{10.3847/1538-3881/ad4a6b}

\bibitem[{{Petrov} \& {Kovalev}(2017{\natexlab{a}})}]{2017MNRAS.467L..71P}
{Petrov}, L., \& {Kovalev}, Y.~Y. 2017{\natexlab{a}}, \mnras, 467, L71,
  \dodoi{10.1093/mnrasl/slx001}

\bibitem[{{Petrov} \& {Kovalev}(2017{\natexlab{b}})}]{r:gaia3}
---. 2017{\natexlab{b}}, \mnras, 471, 3775, \dodoi{10.1093/mnras/stx1747}

\bibitem[{{Petrov} {et~al.}(2011){Petrov}, {Kovalev}, {Fomalont}, \&
  {Gordon}}]{2011AJ....142...35P}
{Petrov}, L., {Kovalev}, Y.~Y., {Fomalont}, E.~B., \& {Gordon}, D. 2011, \aj,
  142, 35, \dodoi{10.1088/0004-6256/142/2/35}

\bibitem[{{Petrov} {et~al.}(2019){Petrov}, {Kovalev}, \&
  {Plavin}}]{2019MNRAS.482.3023P}
{Petrov}, L., {Kovalev}, Y.~Y., \& {Plavin}, A.~V. 2019, \mnras, 482, 3023,
  \dodoi{10.1093/mnras/sty2807}

\bibitem[{{Plank} {et~al.}(2016){Plank}, {Shabala}, {McCallum},
  {Kr{\'a}sn{\'a}}, {Petrachenko}, {Rastorgueva-Foi}, \&
  {Lovell}}]{2016MNRAS.455..343P}
{Plank}, L., {Shabala}, S.~S., {McCallum}, J.~N., {et~al.} 2016, \mnras, 455,
  343, \dodoi{10.1093/mnras/stv2080}

\bibitem[{{Plavin} {et~al.}(2019){Plavin}, {Kovalev}, \&
  {Petrov}}]{2019ApJ...871..143P}
{Plavin}, A.~V., {Kovalev}, Y.~Y., \& {Petrov}, L.~Y. 2019, \apj, 871, 143,
  \dodoi{10.3847/1538-4357/aaf650}

\bibitem[{{Porcas}(2010)}]{2010ivs..conf....8P}
{Porcas}, R. 2010, in Sixth International VLBI Service for Geodesy and
  Astronomy. Proceedings from the 2010 General Meeting, ed. R.~{Navarro},
  S.~{Rogstad}, C.~E. {Goodhart}, E.~{Sigman}, M.~{Soriano}, D.~{Wang}, L.~A.
  {White}, \& C.~S. {Jacobs}, 8--17

\bibitem[{{Reid} \& {Honma}(2014)}]{2014ARA&A..52..339R}
{Reid}, M.~J., \& {Honma}, M. 2014, \araa, 52, 339,
  \dodoi{10.1146/annurev-astro-081913-040006}

\bibitem[{{Rioja} \& {Dodson}(2020)}]{2020A&ARv..28....6R}
{Rioja}, M.~J., \& {Dodson}, R. 2020, \aapr, 28, 6,
  \dodoi{10.1007/s00159-020-00126-z}

\bibitem[{{Schartner} {et~al.}(2023){Schartner}, {Collioud}, {Charlot}, {Xu},
  \& {Soja}}]{2023JGeod..97...17S}
{Schartner}, M., {Collioud}, A., {Charlot}, P., {Xu}, M.~H., \& {Soja}, B.
  2023, Journal of Geodesy, 97, 17, \dodoi{10.1007/s00190-023-01706-4}

\bibitem[{{Schartner} {et~al.}(2024){Schartner}, {Petrachenko}, {Titus},
  {Kr{\'a}sn{\'a}}, {Barrett}, {Hoak}, {Mondal}, {Xu}, \&
  {Soja}}]{2024arXiv240713323S}
{Schartner}, M., {Petrachenko}, B., {Titus}, M., {et~al.} 2024, arXiv e-prints,
  arXiv:2407.13323, \dodoi{10.48550/arXiv.2407.13323}

\bibitem[{{Schuh} \& {Behrend}(2012)}]{2012JGeo...61...68S}
{Schuh}, H., \& {Behrend}, D. 2012, Journal of Geodynamics, 61, 68,
  \dodoi{10.1016/j.jog.2012.07.007}

\bibitem[{{Shepherd} {et~al.}(1994){Shepherd}, {Pearson}, \&
  {Taylor}}]{1994BAAS...26..987S}
{Shepherd}, M.~C., {Pearson}, T.~J., \& {Taylor}, G.~B. 1994, in Bulletin of
  the American Astronomical Society, Vol.~26, 987--989

\bibitem[{{Sokolovsky} {et~al.}(2011){Sokolovsky}, {Kovalev}, {Pushkarev}, \&
  {Lobanov}}]{2011A&A...532A..38S}
{Sokolovsky}, K.~V., {Kovalev}, Y.~Y., {Pushkarev}, A.~B., \& {Lobanov}, A.~P.
  2011, \aap, 532, A38, \dodoi{10.1051/0004-6361/201016072}

\bibitem[{{Titov} {et~al.}(2022){Titov}, {Frey}, {Melnikov}, {Lambert}, {Shu},
  {Xia}, {Gonz{\'a}lez}, {Tercero}, {Gulayev}, {Weston}, \&
  {Natusch}}]{2022MNRAS.512..874T}
{Titov}, O., {Frey}, S., {Melnikov}, A., {et~al.} 2022, \mnras, 512, 874,
  \dodoi{10.1093/mnras/stac038}

\bibitem[{{Xu} {et~al.}(2021{\natexlab{a}}){Xu}, {Anderson}, {Heinkelmann},
  {Lunz}, {Schuh}, \& {Wang}}]{2021JGeod..95...51X}
{Xu}, M.~H., {Anderson}, J.~M., {Heinkelmann}, R., {et~al.} 2021{\natexlab{a}},
  Journal of Geodesy, 95, 51, \dodoi{10.1007/s00190-021-01496-7}

\bibitem[{{Xu} {et~al.}(2017){Xu}, {Heinkelmann}, {Anderson}, {Mora-Diaz},
  {Karbon}, {Schuh}, \& {Wang}}]{2017JGeod..91..767X}
{Xu}, M.~H., {Heinkelmann}, R., {Anderson}, J.~M., {et~al.} 2017, Journal of
  Geodesy, 91, 767, \dodoi{10.1007/s00190-016-0990-x}

\bibitem[{{Xu} {et~al.}(2016){Xu}, {Heinkelmann}, {Anderson}, {Mora-Diaz},
  {Schuh}, \& {Wang}}]{2016AJ....152..151X}
---. 2016, \aj, 152, 151, \dodoi{10.3847/0004-6256/152/5/151}

\bibitem[{{Xu} {et~al.}(2021{\natexlab{b}}){Xu}, {Lunz}, {Anderson},
  {Savolainen}, {Zubko}, \& {Schuh}}]{2021A&A...647A.189X}
{Xu}, M.~H., {Lunz}, S., {Anderson}, J.~M., {et~al.} 2021{\natexlab{b}}, \aap,
  647, A189, \dodoi{10.1051/0004-6361/202040168}

\bibitem[{{Xu} {et~al.}(2022){Xu}, {Savolainen}, {Anderson}, {Kareinen},
  {Zubko}, {Lunz}, \& {Schuh}}]{2022A&A...663A..83X}
{Xu}, M.~H., {Savolainen}, T., {Anderson}, J.~M., {et~al.} 2022, \aap, 663,
  A83, \dodoi{10.1051/0004-6361/202140840}

\bibitem[{{Xu} {et~al.}(2021{\natexlab{c}}){Xu}, {Savolainen}, {Zubko},
  {Poutanen}, {Lunz}, {Schuh}, \& {Wang}}]{2021JGRB..12621238X}
{Xu}, M.~H., {Savolainen}, T., {Zubko}, N., {et~al.} 2021{\natexlab{c}},
  Journal of Geophysical Research (Solid Earth), 126, e2020JB021238,
  \dodoi{10.1029/2020JB02123810.1002/essoar.10504599.1}

\bibitem[{{Xu} {et~al.}(2012){Xu}, {Wang}, \& {Zhao}}]{2012A&A...544A.135X}
{Xu}, M.~H., {Wang}, G.~L., \& {Zhao}, M. 2012, \aap, 544, A135,
  \dodoi{10.1051/0004-6361/201219593}

\bibitem[{{Xu} {et~al.}(2023){Xu}, {Savolainen}, {Bolotin}, {Bernhart},
  {Pl{\"o}tz}, {Haas}, {Varenius}, {Wang}, {McCallum}, {Heinkelmann}, {Lunz},
  {Schuh}, {Zubko}, \& {Kareinen}}]{2023JGRB..12825198X}
{Xu}, M.~H., {Savolainen}, T., {Bolotin}, S., {et~al.} 2023, Journal of
  Geophysical Research (Solid Earth), 128, e2022JB025198,
  \dodoi{10.1029/2022JB02519810.1002/essoar.10512041.1}

\bibitem[{{Xu} {et~al.}(2024){Xu}, {Jung}, {Zhang}, {Xu}, {Byun}, {He},
  {Sakai}, {Titov}, {Shu}, {Kim}, {Cho}, {Yoo}, {Choi}, {Lee}, {Sun}, {Mai}, \&
  {Wang}}]{2024arXiv240907309X}
{Xu}, S., {Jung}, T., {Zhang}, B., {et~al.} 2024, arXiv e-prints,
  arXiv:2409.07309, \dodoi{10.48550/arXiv.2409.07309}

\bibitem[{{Zhang} {et~al.}(2024){Zhang}, {Zhang}, {Xu}, {Liu}, {Chen}, {Ding},
  {Jiang}, {Sun}, {Wang}, {Cui}, {Wen}, {Mai}, {Li}, {Shu}, \&
  {Huang}}]{2024MNRAS.529.2062Z}
{Zhang}, J., {Zhang}, B., {Xu}, S., {et~al.} 2024, \mnras, 529, 2062,
  \dodoi{10.1093/mnras/stae705}

\end{thebibliography}
\bibliographystyle{aasjournal}



\end{document}